\documentclass[twocolumn,pra,showkeys,a4paper]{revtex4-1}
\usepackage[latin1]{inputenc}
\usepackage{amsmath}
\usepackage{mathtools}
\usepackage{amsfonts}
\usepackage{amssymb}
\usepackage{graphicx}
\usepackage[caption=false]{subfig}
\usepackage{tikz}
\usepackage{parskip}
\usepackage{marginnote}
\usetikzlibrary{decorations.pathmorphing}
\usepackage{nomencl}
\makenomenclature

\usepackage{natbib}
\usepackage{booktabs}
\usepackage[export]{adjustbox}
 
\begin{document}
	
	\title{Effect of polydispersity on the phase behavior of additive hard spheres in solution, part I}
	
	\author{Luka Sturtewagen}
	\author{Erik van der Linden}%
	\affiliation{%
	Laboratory of Physics and Physical Chemistry of Foods,\\
	Wageningen University, Bornse Weilanden 9, 6708 WG Wageningen, The Netherlands 
	}%
	
	\begin{abstract}
		We study the theoretical phase behavior of an asymmetric binary mixture of hard spheres, of which the smaller component is monodisperse and the larger component is polydisperse. The interactions are modelled in terms of the second virial coefficient and are assumed to be additive hard sphere interactions. The polydisperse component is subdivided into sub-components and has an average size ten times the size of the monodisperse component. We give the set of equations that defines the phase diagram for mixtures with more than two components in a solvent. We calculate the theoretical liquid-liquid phase separation boundary (the binodal), the critical point and the spinodal. We vary the distribution of the polydisperse component in skewness, modality, polydispersity and number of sub-components. We compare the phase behavior of the polydisperse mixtures with binary monodisperse mixtures for the same average size and binary monodisperse mixtures for the same effective interaction. We find that the largest species in the larger (polydisperse) component causes the largest shift in the position of the phase boundary, critical point and spinodal compared to the binary monodisperse binary mixtures. The polydisperse component also shows fractionation. The smaller species of the polydisperse component favor the phase enriched in the smaller component. This phase also has a higher volume fraction compared to the monodisperse mixture.
	\end{abstract}
	
	\maketitle

	\section{Introduction}
	
	Binary mixtures are usually studied as mixtures of two monodisperse components, however components in nature are usually not that simple, and well defined. Often components will exhibit varying degrees of polydispersity in terms of their size, shape, and charge, which is often ignored when studying phase behavior. 

	Phase separation between hard spheres is driven by two different physical mechanisms. One mechanism involves only excluded volume interactions where the minimum distance between the particles is determined by the sum of their respective radii \cite{Biben1997}, where only a certain asymmetry in the sizes of the particles in the mixture is necessary \cite{Biben1991}. This asymmetry leads to depletion of small spheres around the large spheres and as a result to an effective attraction (depletion interaction) between the larger spheres \cite{Dijkstra1999}.
	This mechanism is referred to as additive hard sphere (HS) model. The case where the minimum distance between the particles is larger or smaller than the sum of their respective radii is referred to as non-additive hard sphere (NAHS) model. In both cases, upon phase separation, the mixture will demix into two (or more) phases, each enriched in one of the components. In this work we will focus on the first type, binary mixtures with significant asymmetry in their size.

	In most studies on the phase behavior of binary mixtures the polydispersity of the components is ignored. However, from experiments with for example gelatin and dextran, it is found that polydispersity has an influence on the phase behavior. The polydispersity of both components leads to significant fractionation, especially for the dextran \cite{Edelman2001}. 

	Polydispersity has an effect on the depletion interaction. With increasing polydispersity, the repulsive barrier decreases, leading to an enhanced rate of flocculation of the large colloidal particles \cite{Goulding2001}. Also Walz \cite{Walz1996} studied the depletion interaction in a solution of normally distributed macromolecules and showed that polydispersity can lead to increased flocculation through the formation of secondary potential energy minima. In later studies they found that polydispersity significantly lowered the magnitude of the repulsive structural barrier, which can be understood in terms of a change in depletion of the macromolecules from the gap \cite{Piech2000}. The polydispersity of the smaller component affects the pair potential between the large particles \cite{Chu1996}. This effect on the depletion interaction can stabilize the particle suspension in the short term, but will still destabilize over time \cite{Tuinier2002}.

	Studying the phase behavior of polydisperse mixtures is challenging, since a polydisperse component effectively consists of a large number of sub-components, each with a different size and possibly also different shape or charge \cite{Sollich2002}. Some theoretical work has been done on predicting the phase behavior of polydisperse components. Cotterman and co-authors used continuous and semi-continuous distributions to predict the fluid-vapor phase diagram of polydisperse components \cite{Cotterman1985} \cite{CottermanChandalla1985}. Santos and co-authors \cite{Santos2010a} studied the phase behavior of polydisperse compounds like polystyrene and polyethylene glycol. They found that the polymer polydispersity played a crucial role in the phase behavior: the broad size distributions lead to a wide range of depletion attractions giving rise to spinodal decomposition preventing gelation. Bellier-Catella and co-authors \cite{Bellier-Castella2000} used a van der Waals approximation for the free energy to study the phase behavior of polydisperse fluids composed of spherical particles. They found the onset of a three phase co-existence at higher polydispersity. Fasolo and co-authors \cite{Fasolo2005} studied theoretically the equilibrium phase behavior of mixtures of polydisperse hard-sphere collids and monodisperse polymers based on the Asakura-Oosawa model. They found that with polydispersity significant fractionation occurred. Polydispersity delayed the onset of both gas-liquid and fluid-solid separation. Also Sear and Frenkel used the Asakura-Oosawa model to predict the phase behavior between a monodisperse colloid and a polydisperse polymer \cite{Sear1997}. They found that polydispersity increased the extent of the fluid-fluid co-existence. Warren and co-authors studied the interaction between hard spheres with a bimodal size distribution and found that demixing caused additional size partitioning and fractionation \cite{Warren1999}. Kang and co-authors used a universal quasichemical (UNIQUAC) model to predict the phase behavior of aqueous polymer systems. They found that the polydispersity of the polymers enlarged the two-phase region considerably near the plait point and resulted in smaller miscibility regions far from the plait point. They also found that the average molecular weights of polymers in the phases differed significantly and this differences increased with larger polydispersity, due to this fractionation, the polydispersity of each polymer was smaller in each child phase compared to the parent mixture \cite{Kang1988}. Others modelled the phase behavior of nondadditive hard-sphere systems using Monte Carlo simulations \cite{Paricaud2008}. They found that with increasing polydispersity the miscibility region decreased and that the critical point shifted towards lower pressures. Also Stapleton and co-authors used Monte Carlo simulations to predict the phase behavior of mixtures with fixed or variable polydispersity \cite{Stapleton1990}. They found that mixtures even with a very small degree of polydispersity resulted in differences in the phase separation and the fractionation between the coexisting phases. 

	In this study we aim to get a better understanding of how size polydispersity influences the phase behavior in binary mixtures, mainly on the position of the phase boundary, the spinodal, and the critical point. Next to that we aim to predict the fractionation of the polydisperse component between the phases. We model the interactions between the different components using the second virial coefficient. 
	We start the theoretical considerations with reviewing the interaction in a simple system of a solute in a solvent (section \ref{ss:dilute}). In section \ref{ss:singlesolute} and \ref{ss:multiple} we expand the second virial coefficients for solutions with one type of solute component to solutions with multiple distinguishable types of solute components. Section \ref{ss:stability} describes the theory about the stability of a mixture, section \ref{ss:critical} describes the theory about the critical point and finally section \ref{ss:phaseboundary} describes the theory about the phase boundary. We chose to first describe the existing theory in order to more easily explain the expressions we used in our calculations. With the expressions in \ref{s:Theory} we calculate the phase behavior for different mixtures with varying polydispersity. 
	In section \ref{s:RD} we discuss the resulting phase diagrams, first in \ref{ss:2pseudo} we divided the polydisperse component into two sub-components and in section \ref{ss:9pseudo} we increased the number of sub-components to nine. Finally we discuss fractionation of the polydisperse component in section \ref{ss:Fraction}.
	\section{\label{s:Theory}Theory}
	We start by deriving the equations of state for dilute solutions. Next we derive the virial expansion for solutions with one solute component. Subsequently we derive the second virial coefficient for mixtures with an aribtrary number of distinguishable components. This gives us all the parameters we need to define the stability boundary, the critical point and phase boundary of a mixture. The resulting system of equations is solved in Matlab R2017b.
	\subsection{\label{ss:dilute}Dilute liquid solutions}
	We consider a two component solution, in which one component is the solvent and the other component is the solute. We define $N_s$ as the number of solvent particles and $N_{\nu}$  as the number of solute particles in a volume $V$ at a temperature $T$. The total number of particles in the system is then $N = N_{\nu}+N_s$ and since we assume a dilute solution, $N_s >> N_{\nu}$. The system is in constant thermal contact with the environment and both the volume, and the number of solute and solvent particles are fixed (canonical ensemble) \cite{Hill1986}.
	
	The sum of the kinetic $(K)$ and potential energies $(U)$ of the system represents the Hamiltonian ($H$) of the system, given by 
	\begin{equation}
	H(p^{N_{\nu}+N_s},q^{N_{\nu}+N_s}) = K(p^{N_{\nu}+N_s}) + U(r^{N_{\nu}+N_s}) \label{eq:Hamiltonian}
	\end{equation}
	in which 
	\begin{align}
	K(\vec{p}_1...\vec{p}_{N_{\nu}+N_s}) &= \sum_{i=1}^{N_{\nu}} \frac{|\vec{p}_i|^2}{2m}+\sum_{i=N_{\nu}+1}^{N_{\nu}+N_s}\frac{|\vec{p}_i|^2}{2m_s} \label{eq:Kinetic}\\
	U(\vec{r}_1...\vec{r}_{N_{\nu}+N_s}) &=  \sum_{i < j}^{N_{\nu}+N_s}\phi_{ij}(\vec{r}_i-\vec{r}_j)\nonumber\\
	&= \sum_{i < j}^{N_{\nu}}\phi_{ij}(\vec{r}_i-\vec{r}_j) + \sum_{i < j}^{N_s}\phi_{ij}(\vec{r}_i-\vec{r}_j)\nonumber\\
	& \quad+ \sum_{i}^{N_{\nu}}\sum_{j}^{N_s}\phi_{ij}(\vec{r}_i-\vec{r}_j) \nonumber\\
	&= U^{N_{\nu}}+U^{N_s}+U^{N_{\nu}N_s} \label{eq:Potential}
	\end{align}
	\nomenclature{$H$}{Hamiltonian}
	\nomenclature{$p^{i}$}{Momentum of particle $i$}
	\nomenclature{$q^{i}$}{Velocity of particle $i$}
	\nomenclature{$K$}{Kinetic energy}
	\nomenclature{$U$}{Potential energy}
	\nomenclature{$\vec{r}_i$}{Position of the particle}
	\nomenclature{$\vec{p}_i$}{Impulse of the particle}
	\nomenclature{$\phi_{ij}$}{Pair potential between particle $i$ and $j$}
	\nomenclature{$N_{\nu}$}{Number of solute particle}
	\nomenclature{$m_{\nu}$}{Mass of the solute particle}
	\nomenclature{$m_s$}{Mass of the solvent particle}
	\nomenclature{$N_s$}{Number of solvent particles}
	\nomenclature{$N$}{Number of particles}
	\nomenclature{$T$}{Temperature}
	\nomenclature{$V$}{Volume}
	
	where $\vec{p_i}$ is the impulse of particle $i$, $m_{\nu}$ is the mass of a solute particle and $m_s$ is the mass of a solvent particle, $\phi_{ij}$ is the pair potential between particle $i$ and $j$ and $\vec{r_i}$ is the position of particle $i$.
	
	The canonical partition function $(Z)$ describes the statistical properties of the system for a given temperature, volume, and number of particles. The partition function is the sum of all the different individual energy states in which the system can exist. The states of the system are specified by both the position and the momentum of the particles. Applying the partition function to dilute solutions makes it possible to reduce the many-body problem in statistical mechanics to problems of one-body, two-body, three-body, etc.
	
	\begin{multline}
	Z(T,V,N_{\nu},N_s)= \frac{h^{-(3N_{\nu}+3N_s)}}{N_{\nu}!N_s!}\\ 
	\times \int_V \exp{(-\beta  H(p^{N_{\nu}},p^{N_s},q^{N_{\nu}},q^{N_s}))} \\ 
	\times d \vec{r}_1...d\vec{r}_{N_{\nu}+N_s}d  \vec{p}_1...d\vec{p}_{N_{\nu}+N_s}\label{eq:Partition}
	\end{multline}
	\nomenclature{$h$}{Planck's constant}
	\nomenclature{$k$}{Boltzmann's constant}
	\nomenclature{$\beta$}{$\displaystyle\frac{1}{kT}$}
	\nomenclature{$Z$}{Canonical partition function}
	where $h$ is Plank's constant and $\displaystyle\beta = \frac{1}{kT}$, in which $k$ is Boltzmann's constant. 
	
	Other thermodynamic variables, such as the Helmholtz free energy, the pressure and the chemical potential can be expressed in terms of this function or its derivatives. The Helmholtz free energy $(A)$ for this system is then given by \cite{Hill1986}:
	\begin{equation}
	A(T,V,N_{\nu},N_s) = -kT \ln(Z(T,V,N_{\nu},N_s)) \label{eq:Helmholtz}
	\end{equation}
	With the differential of the free energy given by:
	\begin{equation}
	d A = -S d T - p d V +\sum_i \mu_i d N_i \label{eq:dHelmholtz}
	\end{equation}
	\nomenclature{$A$}{Helmholtz free energy}
	Therefore, the pressure ($p$) is given by:
	\begin{equation}
	p = -\left(\frac{\partial A}{\partial V}\right)_{T,N} = kT\left(\frac{\partial Z}{\partial V}\right)_{T,N} \label{eq:Pressure}
	\end{equation}
	\nomenclature{$p$}{The pressure}
	and the chemical potential ($\mu_{i}$) for component $i$ is given by:
	\begin{equation}
	\mu_i = -\left(\frac{\partial A}{\partial N_i}\right)_{T,V,N_{\alpha \neq i}} = kT\left(\frac{\partial Z}{\partial N_i}\right)_{T,V,N_{\alpha \neq i}} \label{eq:Chempot}
	\end{equation}
	Since we focus on particles with hard sphere interaction, we can integrate out the momentum integrals in eq. \ref{eq:Partition}.
	\begin{multline}
	Z(T,V,N_{\nu},N_s)=\frac{\Lambda_{\nu}^{-3N_{\nu}}\Lambda_s^{-3N_s}}{N_{\nu}!N_s!}\\
	\times \int_V d \vec{r}_1...d\vec{r}_{N_{\nu}+N_s} \exp{(-\beta U(\vec{r}_1...\vec{r}_{N_{\nu}+N_s}))} \\
	= \Lambda_{\nu}^{-3N_{\nu}}\Lambda_s^{-3N_s}Q(T,V,N_{\nu},N_s) \label{eq:Partition1}
	\end{multline}
	\nomenclature{$\Lambda$}{The thermal wavelegnth $\displaystyle\left(\frac{h^2}{2\pi m k T}\right)^{1/2}$}
	where $\displaystyle\Lambda = \left(\frac{h^2}{2\pi m k T}\right)^{1/2}$ is the mean thermal wavelength and $Q$ the configuration integral. The configuration integral is the integral over all possible configurations of the $N$ molecules in the system:
	\begin{multline}
	Q(T,V,N_{\nu},N_s) = \frac{1}{N_{\nu}!N_s!}\int_V d \vec{r}_1...d\vec{r}_{N_{\nu}+N_s}\\
	\times \exp{(-\beta U(\vec{r}_1,...,\vec{r}_{N_{\nu}+N_s}))} \label{eq:ConfigIntegral}
	\end{multline}
	\nomenclature{$Q$}{Configuration integral}
	
	The first three configuration integrals are:
	\begin{align}
	&Q_1 = \int_{V}d\vec{r}_1 = V \label{eq:Q1} \\
	&Q_2 = \frac{1}{2} \int_{V}d\vec{r}_1 d\vec{r}_2 \exp\left[-\beta \phi (|\vec{r}_1-\vec{r}_2|)\right] \label{eq:Q2} \\
	&Q_3 = \frac{1}{6} \int_{V}d\vec{r}_1 d\vec{r}_2 d\vec{r}_3\nonumber\\
	&\quad\times \exp \left\lbrace -\beta \left[\phi(|\vec{r}_1-\vec{r}_2|) + \phi(|\vec{r}_1-\vec{r}_3|) + \phi(|\vec{r}_2-\vec{r}_3|)\right]\right\rbrace \label{eq:Q3}
	\end{align}
	The configuration integral $Q_1$ indicates there is only one particle present in our selected volume. $Q_2$ indicates there are two particles present in our selected volume: interacting or not interaction. $Q_3$ indicates there are three particles present in the volume. These particles can interact with each other or not interact. The number of combinations of interactions increases significantly when increasing the number of particles. 
	
	The configuration integrals can be represented in diagrams where each dot is a particle present in the volume and a line between the dots indicates interaction:
	\begin{align}
	Q_1 &=  \tikz{\fill[black] (0,0) circle (.1 cm)} \label{eq:Q1dot}\\
	Q_2 &= \frac{1}{2}\left( \tikz{\fill[black] (0,0) circle (.1 cm); \fill[black] (0.5,0) circle (.1 cm)} +  \tikz{\fill[black] (0,0) circle (.1 cm); \fill[black] (0.5,0) circle (.1 cm); \draw(0,0) -- (.5,0)} \right) \label{eq:Q2dot} \\
	Q_3 &= \frac{1}{6}\left( \tikz{\fill[black] (0,0) circle (.1 cm); \fill[black] (0.5,0) circle (.1 cm); \fill[black] (0.25,0.43) circle (.1 cm)} + 3 \tikz{\fill[black] (0,0) circle (.1 cm); \fill[black] (0.5,0) circle (.1 cm); \draw(0,0) -- (.5,0); \fill[black] (0.25,0.43) circle (.1 cm)} + 3 \tikz{\fill[black] (0,0) circle (.1 cm); \fill[black] (0.5,0) circle (.1 cm); \draw(0,0) -- (.5,0); \fill[black] (0.25,0.43) circle (.1 cm);\draw(0,0) -- (.25,.43)} + \tikz{\fill[black] (0,0) circle (.1 cm); \fill[black] (0.5,0) circle (.1 cm); \draw(0,0) -- (.5,0); \fill[black] (0.25,0.43) circle (.1 cm);\draw(0,0) -- (.25,.43);\draw(0.5,0) -- (.25,.43)}\right) \label{eq:Q3dot}
	\end{align}
	
	In our analysis we consider solutions where the solvent particles are present in a much larger number than the solute particles. This means that solute particles have relatively low influence on the statistics of the solvent particles. Following the MacMillan - Mayer theory we can describe the interactions between the solute particles by a potential of mean force equation and thus can apply additivity of the particle-particle interactions in eq. \ref{eq:Potential} \citep{McMillan1945}.
	\begin{multline}
	U(\vec{r}_1,...,\vec{r}_{N_{\nu}+N_s}) \\
	= U_{N_{\nu}N_{\nu}}(\vec{r}_1,...,\vec{r}_{N_{\nu}}) + U_{N_{\nu}N_s}(\vec{r}_1,...,\vec{r}_{N_{\nu}+N_s}) \\ + U_{N_sN_s}(\vec{r}_{N_{\nu}+1},...,\vec{r}_{{N_{\nu}}+N_s}) \label{eq:PotentialSum}
	\end{multline}
	The potential of mean force for dissolved particles ($W$) is defined according to:
	\begin{multline}
	\exp\left(-\beta W(\vec{r}_1,...,\vec{r}_{N_\nu})\right) =\\ \dfrac{\int d\vec{r}_{N_{\nu}+1},...,d\vec{r}_{N_{\nu}+N_s}\exp\left(-\beta U(\vec{r}_1,...,\vec{r}_{N_{\nu}+N_s})\right)} {\left(\splitdfrac{\int d\vec{r}_{N_{\nu}+1},...,d\vec{r}_{N_{\nu}+N_s}}{\times\exp\left(-\beta U_{N_sN_s}(\vec{r}_{N_{\nu}+1},...,\vec{r}_{N_{\nu}+N_s}) \right)}\right)}\\ \label{eq:MeanForce}
	\end{multline}
	\nomenclature{$W_{\nu}$}{Pair potential for disolved compounds}
	Using eq. \ref{eq:ConfigIntegral} and eq. \ref{eq:MeanForce}, the configuration integral becomes \cite{Vafaei2014}:
	\begin{multline}
	Q(T,V,N_{\nu},N_s) = \\ \frac{1}{N_{\nu}!} \int d \vec{r}_1...d \vec{r}_N \exp\left[-\beta W_{\nu}(\vec{r}_1,...,\vec{r}_{N_{\nu}})\right] 
	\\ \times \frac{1}{N_s!} \int d \vec{r}_{N_{\nu}+1} ... d \vec{r}_{N_{\nu}+N_s}\\\times \exp \left[-\beta U_{N_sN_s}(\vec{r}_{N+1},...,\vec{r}_{N+N_s})\right] \\ = Q_W(T,V,N_{\nu},\mu_s)Q_s(T,V,N_s) \label{eq:Conf}
	\end{multline}
	The Helmholtz free energy (eq. \ref{eq:Helmholtz}) of the system then becomes the sum of the Helmholtz free energy of the solvent and the Helmholtz free energy of the solute:
	\begin{multline}
	A(T,V,N_{\nu},N_S)\\ = -kT \ln(Z_W(T,V,N_{\nu},\mu_s)) - kT \ln(Z_s(T,V,N_s))\label{eq:Helmholtzsum}
	\end{multline}
	\subsection{\label{ss:singlesolute}Second virial coefficient of a dilute solution with a single solute component}
	Similar to the expansion of the universal gas law by a virial expansion for real gasses, we can write a virial expansion for the osmotic pressure, $\Pi$, of a solution according to:
	\begin{equation}
	\beta\Pi = \rho + B_2(T,\mu_s)\rho^2 + B_3(T,\mu_s)\rho^3 + ... \label{eq:EquationofState}
	\end{equation}
	\nomenclature{$\Pi$}{Osmotic pressure}
	\nomenclature{$\rho$}{Density $\displaystyle\frac{N}{V}$}
	\nomenclature{$B$}{Virial coefficient}
	\nomenclature{$B_2$}{Second virial coefficient}
	with $\rho$ the number density of the component $\displaystyle\left(\frac{N_\nu}{V}\right)$, $B_2$ the second virial coefficient, and $B_3$ the third virial coefficient. The second virial coefficient accounts for the increase in osmotic pressure due to particle pairwise interaction. The third virial coefficient accounts for the interaction between three particles. The equation can be expanded for higher densities with $B_n$, the $n^{th}$ virial coefficient, which accounts for the interaction between $n$ different particles.
	
	Until now we have been using the canonical ensemble to describe the system. In the canonical ensemble, the number of particles ($N_\nu+N_s$), the temperature ($T$), and the volume ($V$) are fixed. The restraint of constant number of particles becomes tedious when accounting for the interaction between the different particles, therefore we will use the grand canonical ensemble to derive the virial coefficients. In the grand canonical ensemble, the temperature ($T$) and the volume ($V$) are fixed, as well as the chemical potentials $(\mu_i)$.
	
	We can write the grand canonical partition function independent on $N_{\nu}$ and $N_s$ by performing a transformation of the number of compounds by the chemical potential $\mu_{\nu}$ and $\mu_s$ \citep{Hill1986}.
	\begin{align}
	\Xi(T,V,\mu_{\nu},\mu_s) &= \sum_{N_{\nu}=0}^{\infty} e^{\beta\mu_{\nu}N_{\nu}}\nonumber\\
	&\quad\quad\times \sum_{N_s=0}^\infty e^{\beta\mu_sN_s} Z(T,V,N_{\nu},N_s) \nonumber\\
	& = \sum_{N_{\nu}=0}^\infty e^{\beta\mu_{\nu}N_{\nu}}Z_W(T,V,N_\nu,\mu_s)\nonumber\\
	&\quad\times\Xi_s(T,V,\mu_s)\nonumber\\
	& = \Xi_W(T,V,\mu_{\nu},\mu_s)\Xi_s(T,V,\mu_s) \label{eq:GrandPartFunction}
	\end{align}
	\nomenclature{$\Xi$}{Grand canonical partition function}
	\nomenclature{$\mu$}{Chemical potential}
	The equation of state for the system is given by:
	\begin{multline}
	\beta(p + \Pi) =  \frac{\ln{(\Xi(T,V,\mu_{\nu},\mu_s))}}{V}\\ = \frac{\ln{(\Xi_s(T,V,\mu_s))}}{V} +  \frac{\ln{(\Xi_W(T,V,\mu_{\nu},\mu_s))}}{V}
	\end{multline}
	In which the osmotic pressure is given by:
	\begin{equation}
	\beta\Pi = \frac{\ln{(\Xi_W(T,V,\mu_{\nu},\mu_s))}}{V} \label{eq:OsmPressure}
	\end{equation}
	\nomenclature{$z$}{Activity $\displaystyle\frac{e^{\beta\mu}}{\Lambda^3}$}
	We can now define the activity as $\displaystyle z\equiv\frac{e^{\beta\mu}}{\Lambda^3}$
	\begin{multline}
	\Xi_W(T,V,z_{\nu},\mu_s) = \sum_{N=0}^\infty Q_N(T,V,N_\nu,\mu_s)z_{\nu}^N\\
	=1+Q_1z+Q_2z^2+Q_3z^3+... \label{eq:GrandPartAct}
	\end{multline}
	
	The osmotic pressure can be written in terms of the logarithm of the grand canonical ensemble.
	\begin{align}
	\beta\Pi(T,V,z_{\nu},\mu_s)V &= \ln\left(\Xi_W(T,V,z_{\nu},\mu_S)\right) \nonumber\\
	&= \ln(1+Q_1z+Q_2z^2+Q_3z^3+...) \nonumber\\
	&=Q_1z+\left(Q_2-\frac{1}{2}Q^2_1\right)z^2\nonumber\\
	&\quad+\left(Q_3-Q_1Q_2+\frac{1}{3}Q^3_1\right)z^3+... \label{eq:OsmoticPressureQ}
	\end{align}
	This can be written as:
	\begin{equation}
	\beta \Pi = \sum_{l=1}^{\infty} b_l z^l \label{eq:OsmoticPressureCluster}
	\end{equation}
	The coefficient $b_l$ is also known as the cluster-integral and indicate interaction among $l$ compounds.
	\begin{align}
	b_1 &= \frac{Q_1}{V} \label{eq:cluster1}\\
	b_2 &= \frac{Q_2-\frac{1}{2}Q^2_1}{V} \label{eq:cluster2}\\
	b_3 &= \frac{Q_3-Q_1Q_2+\frac{1}{3}Q^3_1}{V} \label{eq:cluster3}
	\end{align}
	Just as that configuration integrals can be written in diagrams (see eq. \ref{eq:Q1dot} - \ref{eq:Q3dot}), the cluster integrals can also be written as diagrams \cite{Hansen2012}: 
	\begin{align}
	b_1 &=  \frac{1}{V}\left(\tikz{\fill[black] (0,0) circle (.1 cm)}\right) \label{eq:cluster1dot}\\
	b_2 &= \frac{1}{2V}\left(\tikz{\fill[black] (0,0) circle (.1 cm); \fill[black] (0.5,0) circle (.1 cm); \draw(0,0) -- (.5,0)} \right) \label{eq:cluster2dot}\\
	b_3 &= \frac{1}{6V}\left(3 \tikz{\fill[black] (0,0) circle (.1 cm); \fill[black] (0.5,0) circle (.1 cm); \draw(0,0) -- (.5,0); \fill[black] (0.25,0.43) circle (.1 cm);\draw(0,0) -- (.25,.43)} + \tikz{\fill[black] (0,0) circle (.1 cm); \fill[black] (0.5,0) circle (.1 cm); \draw(0,0) -- (.5,0); \fill[black] (0.25,0.43) circle (.1 cm);\draw(0,0) -- (.25,.43);\draw(0.5,0) -- (.25,.43)}\right) \label{eq:cluster3dot}
	\end{align}
	\nomenclature{$b_l$}{Clusterintegral of $l$ particles}
	\nomenclature{$f_{12}$}{Mayer f-function for particle 1 and 2}
	Two particle interactions can be defined by the Mayer f-function:  $$\displaystyle f_{12} = \exp(-\beta W_{12})-1$$
	and can be represented by:
	\begin{center}
		\begin{tikzpicture}
		\draw(0.2,0) -- (.8,0);
		\draw (0,0) circle (.2 cm) node {1};
		\draw (1,0) circle (.2 cm) node {2};
		\end{tikzpicture}
	\end{center}	
	Substituting the Mayer f-function into the cluster integrals we get:
	\begin{align}
	b_1 &= \frac{1}{V} \int_{V}d\vec{r}_1 = 1 \label{eq:cluster1Mayer}\\
	b_2 &=  \frac{1}{2V} \int_{V}d \vec{r}_1 d\vec{r}_2 f_{12} = 2\pi\int_{0}^{\infty}dr r^2
	f(r) \label{eq:cluster2Mayer}\\
	b_3 &= 2 b^2_2 + \frac{1}{6}\int d\vec{r} \int d\vec{r}' f(r) f(r') f(|\vec{r}-\vec{r}'|) \label{eq:cluster3Mayer} 
	\end{align}
	In order to find the relationship between cluster integrals and the virial coeffcients, we need to do several substitutions and inversions using also eq. \ref{eq:OsmoticPressureCluster}:
	\begin{multline}
	\rho=\frac{N}{V}=\beta z \left(\frac{\partial \Pi}{\partial z}\right)T \\
	= \sum_{l=1}^\infty l b_l z^l = z + 2b_2z^2 + 3b_3z^3+... \label{eq:density}
	\end{multline}
	Inverting the series we obtain for the activity \cite{abramowitz2012handbook}:
	\begin{equation}
	z= \rho -2b_2 \rho^2 + (8b^2_2-3b_3)\rho^3+... \label{eq:activity}
	\end{equation}
	This can be substituted in the equation for the pressure (eq. \ref{eq:OsmoticPressureCluster}):
	\begin{equation}
	\beta \Pi = \rho - b_2 \rho^2 + 2(2 b^2_2-b_3)\rho^3 + ... \label{eq:OsmoticPressureseries}
	\end{equation}
	Comparing this equation to eq. \ref{eq:EquationofState}, we see that the second virial coefficient is equivalent to $-b_2$:
	\begin{align}
	B(T,\mu_s) &= -b_2 = -\frac{1}{2V}\left(\tikz{\fill[black] (0,0) circle (.1 cm); \fill[black] (0.5,0) circle (.1 cm); \draw(0,0) -- (.5,0)} \right)\nonumber\\
	&= 2 \pi \int_{0}^{\infty}d r r^2(1-\exp{(-\beta W(r)))}\nonumber\\
	&= -2\pi\int_{0}^{\infty}dr r^2 f(r) \label{eq:SecondVirialCoeff}
	\end{align}
	For hard spheres we define the interaction potentials as for components with a diameter $\sigma$: 
	\begin{equation}
	 W(r)_{HS} = \left\{
	 \begin{array}{ll} 0, & r > \sigma \\
	 \infty, & r \leq \sigma
	 \end{array}\right. \label{eq:HSpotential}
	\end{equation}
	\nomenclature{$\sigma$}{Diameter of a particle}
	
	The second virial coefficient then becomes:
	\begin{equation}
	B(T,\mu_s) = \frac{2\pi}{3}(\sigma)^3 \label{eq:SecondVirialHS}
	\end{equation}
	\subsection{\label{ss:multiple}Second virial coefficient of solutions with multiple solute compounds}
	When there are more distinct compounds ($\nu_1,..,\nu_n$) in the solution with $n$ the total number of distinguishable compounds, or species, there are two main types of two particle interactions that can occur: 
	\begin{itemize}
		\item interactions between indistinguishable components, i.e. components of the same species:
		\begin{itemize}
			\item \tikz{\draw(0,0) -- (.5,0);\fill[blue] (0,0) circle (.1cm);\fill[blue] (.5,0) circle (.1cm)}, \tikz{\draw(0,0) -- (.5,0);\fill[red] (0,0) circle (.1cm);\fill[red] (.5,0) circle (.1cm)}, \tikz{\draw(0,0) -- (.5,0);\fill[green] (0,0) circle (.1cm);\fill[green] (.5,0) circle (.1cm)}, \tikz{\draw(0,0) -- (.5,0);\fill[orange] (0,0) circle (.1cm);\fill[orange] (.5,0) circle (.1cm)},...
		\end{itemize}
		\item interactions between distinguishable components, i.e. components of different species:
		\begin{itemize}
			\item \tikz{\draw(0,0) -- (.5,0);\fill[blue] (0,0) circle (.1cm);\fill[red] (.5,0) circle (.1cm)}, \tikz{\draw(0,0) -- (.5,0);\fill[green] (0,0) circle (.1cm);\fill[orange] (.5,0) circle (.1cm)}, \tikz{\draw(0,0) -- (.5,0);\fill[blue] (0,0) circle (.1cm);\fill[green] (.5,0) circle (.1cm)}, \tikz{\draw(0,0) -- (.5,0);\fill[red] (0,0) circle (.1cm);\fill[orange] (.5,0) circle (.1cm)},...
		\end{itemize}
	\end{itemize}
	\nomenclature{$\nu_i$}{Compound of type $\nu_i$}
	\nomenclature{$n$}{Number of distinguishable components}
	We can write the configuration integral $Q_N$ in general as:
	\begin{multline}
	Q_N = \frac{1}{N_{\nu_1}!...N_{\nu_n}!}\\ \times\int_{V}d\vec{r}_{1_{\nu_1}},...,\vec{r}_{N_{\nu_1}},...,\vec{r}_{1_{\nu_n}},..,\vec{r}_{N_{\nu_n}}\\
	\times\exp\left[-\beta\sum_{i<j}^{N}W_{xy}(r_{ij})\right] \label{eq:GeneralConfiguration}
	\end{multline} 
	In which: $\displaystyle N = \sum_{i}^{n} N_{\nu_i}$ or the total number of particles in the configuration and $x$ and $y$ can be of any type $\nu_n$ in the mixture.
	
	The general equation for the partition function in the grand canonical ensemble then becomes:
	\begin{multline}
	\Xi(T,V,z_{\nu_1},..., z_{\nu_n},\mu_s) =\\
	\sum_{N_{\nu_1},..., N_{\nu_n} \geqslant 0} Q_{N}(V,T) z_{\nu_1}^{N_{\nu_1}} ... z_{\nu_n}^{N_{\nu_n}} \label{eq:GeneralPartitionFunction}
	\end{multline}
	In the case of two particle interaction we have interaction between components of the same species and interaction between components of different species, so we obtain for $Q_2$ two types of configuration integrals (comparable to eq. \ref{eq:Q2}):
	\begin{align}
	Q_{2,xx} &= \frac{1}{2!} \int_{V}d\vec{r}_{1_x} d \vec{r}_{2_x} \exp{\left[-\beta W_{xx}(r_{1_x} r_{2_x})\right]} \nonumber\\
	&= \frac{1}{2}V^2+2 \pi V\int_{0}^{\infty}dr r^2f_x(r) \label{eq:Qxx} \\
	Q_{2,xy} &= \frac{1}{1!1!} \int_{V}d\vec{r}_{1_x} d\vec{r}_{2_y} \exp{\left[-\beta W_{xy}(r_{1_x} r_{2_y})\right]} \nonumber\\
	&= 2\left(\frac{1}{2}V^2 + 2 \pi V\int_{0}^{\infty}dr r^2 f_{xy}(r)\right) \label{eq:Qxy}
	\end{align}
	In which $x$ and $y$ can be of any type $\nu_n$ in the mixture and $y \neq x$.
	
	Also for the cluster integrals we obtain two types of integrals (comparable to eq. \ref{eq:cluster2Mayer}):
	\begin{align}
	b_{2,xx} &= 2\pi \int_{0}^{\infty} dr r^2\left(\exp{\left[-\beta W_{xx}\right]}-1\right) \nonumber\\
	&= 2\pi \int_{0}^{\infty} dr r^2 f_x(r) \label{eq:clusterxx}\\
	b_{2,xy} &= 4\pi \int_{0}^{\infty} dr r^2\left(\exp{\left[-\beta W_{xy}\right]}-1\right)\nonumber\\
	&= 2\left(2\pi \int_{0}^{\infty} dr r^2 f_{xy}(r)\right) \label{eq:clusterxy}
	\end{align}
	In which $x$ and $y$ can be of any type $\nu_n$ in the mixture and $y \neq x$.
	
	The general equation for the osmotic pressure becomes then (comparable to eq. \ref{eq:density}):
	\begin{align}
	\beta\Pi &= b_{1,\nu_1}z_{\nu_1} + b_{1,\nu_2}z_{\nu_2} + b_{1,\nu_3}z_{\nu_3} + ... \nonumber\\
	&+ b_{2,\nu_1\nu_1}z_{\nu_1}^2 + b_{2,\nu_1\nu_2}z_{\nu_1} z_{\nu_2} + b_{2,\nu_1\nu_3}z_{\nu_1} z_{\nu_3} + ... \nonumber\\
	&= \sum_{i}^{n}b_{1,\nu_i}z{\nu_i} + \sum_{i}^{n}b_{2,\nu_i\nu_i}z_{\nu_i}^2 + \sum_{i<j}^{n}b_{2,\nu_i\nu_j}z_{\nu_i}z_{\nu_j} + ... \label{eq:GeneralOsmotic}
	\end{align}
	For the second virial coefficient we obtain two types (comparable to eq. \ref{eq:SecondVirialCoeff}):
	\begin{align}
	B_{xx}^* &= 2 \pi \int_{0}^{\infty} dr r^2(1-\exp\left[-\beta W_{xx}\right])\nonumber\\
	&= -2 \pi \int_{0}^{\infty}dr r^2f_x(r) = B_{xx} \label{eq:Bxx}\\
	B_{xy}^* & =4 \pi \int_{0}^{\infty} dr r^2(1-\exp\left[-\beta W_{xy}\right]) \nonumber\\
	&= 2 \left(-2 \pi \int_{0}^{\infty}dr r^2f_{xy}(r) \right) = 2 B_{xy} \label{eq:Bxy}
	\end{align}
	In which $x$ and $y$ can be of any type $\nu_n$ in the mixture and $y \neq x$. Note: we define a $B^*$ to have all the second virial equations of the same form: $-2\pi\int_{0}^\infty dr r^2f(r)$, with $f(r)$ dependent on the type of interaction.
	
	For additive hard sphere interaction, the interaction potential for particles of different species is given by:
	\begin{equation}
		W(r)_{HS} = \left\{
		\begin{array}{ll} 0, & r > \sigma_{ij} \\
		\infty, & r \leq \sigma_{ij}
		\end{array}\right. \label{eq:HScrosspotential}
	\end{equation}
	with $\sigma_{ij} = (\sigma_i + \sigma_j)/2$, the distance between the centers of the two components. When the interaction is not additive, the distance of closest approach of the centers of the two components becomes: $\sigma_{ij} = 1/2(\sigma_i + \sigma_j)(1+\Delta)$, in which $\Delta$ accounts for the non-additivity of the interaction between the particles that are different.
	
	For the second virial coefficient we find (comparable to eq. \ref{eq:SecondVirialHS}):
	\begin{align}
	B_{xx} &= \frac{2\pi}{3}(\sigma_x)^3 \label{eq:B_xx}\\
	B_{xy} &= \frac{2\pi}{3}\left(\left(\frac{\sigma_x+\sigma_y}{2}\right)(1+\Delta)\right)^3 \label{eq:B_xy}
	\end{align}
	\nomenclature{$\Delta$}{Non-additivity parameter for interacting particles}
	(Again, for additive hard sphere interactions, $\Delta = 0$).
		
	The general equation for the osmotic pressure for a dilute mixture is then given by:
	\begin{align}
	\beta \Pi &= \rho + B_{\nu_1\nu_1}\rho_{\nu_1}^2 + 2B_{\nu_1\nu_2}\rho_{\nu_1} \rho_{\nu_2} + 2B_{\nu_1\nu_3}\rho_{\nu_1} \rho_{\nu_3}  ... \nonumber\\
	&= \rho + \sum_{i}^{n}\sum_{j}^{n}B_{\nu_i\nu_j}\rho_{\nu_i}\rho_{\nu_j} +... \label{eq:OsmoticGeneral}
	\end{align}
	\nomenclature{$x$}{Particle of type $x$}
	\nomenclature{$y$}{Particle of type $y$}
	The second virial coefficients can be represented in matrix form:
	\begin{equation}
	B = \begin{bmatrix}
	B_{11} & \cdots & B_{1n} \\
	\vdots & \ddots & \vdots \\
	B_{1n} & \cdots & B_{nn} \label{eq:VirialMatrix}
	\end{bmatrix}
	\end{equation}
	In which we abbreviate the notation $B_{\nu_i\nu_j}$ to $B_{ij}$, and similarly, the densities $\rho_{\nu_i}$ by $\rho_i$.
	\subsection{\label{ss:stability}Stability of a mixture}
	The stability of a mixture is dependent on the second derivative of the free energy. If the second derivative of the mixture becomes zero, the mixture is at the boundary of becoming unstable. Unstable mixtures have a negative second derivative  \cite{Beegle1974} \cite{Heidemann1975}.
	
	The free energy of a mixture is given by \cite{Hill1986}:
	\begin{multline}
	A(T,V,N_{\nu_1},..., N_{\nu_n},\mu_s) \\= -kT \ln(Z(T,V,N_{\nu_1},..., N_{\nu_n},\mu_s)) \label{eq:Helmholtzgeneral}
	\end{multline}
	and the differential is given by:
	\begin{equation}
	d A = -S d T - p dV + \sum_i^n \mu_{i} d N_{i} \label{eq:Helmholtzdifferential}
	\end{equation}
	\nomenclature{$S$}{Entropy}
	in which the chemical potential (the first partial derivative of the free energy with respect to number of particles ($N_i$)) for component $i$ is given by:
	\begin{equation}
	\mu_i = \mu_i^0 + kT \ln(\rho_i)+2kT\left(\sum_{j}^{n} B_{ij}\rho_j\right) \label{eq:Chempotential}
	\end{equation}
	
	\begin{figure*}
		\centering
		\includegraphics[trim=0cm .5cm 0cm 1cm, clip=true ,width=0.78\textwidth]{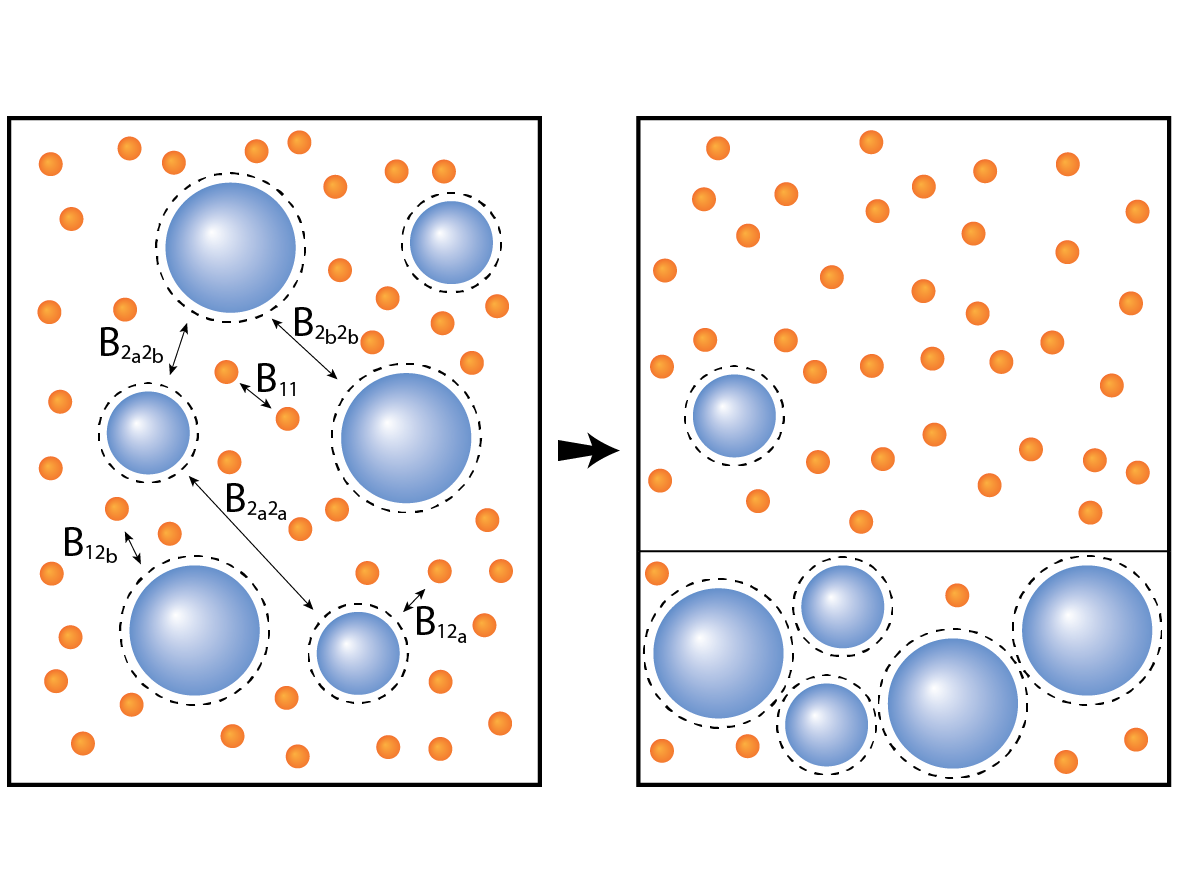}
		\caption{Graphical representation of a simple polydisperse mixture, in which the polydisperse component consists out of two-sub components ($a$ and $b$, $n=3$), second virial coefficients are indicated. The mixture demixes into two phases, one phase enriched in the small component, one phase enriched in the large polydipserse component}
		\label{fig:GraphicalDiagram}
	\end{figure*}
	
	For a mixture with $n$ distinguishable components, this second partial derivatives can be represented by a $n\times n$ matrix of the first partial derivatives of the chemical potential of each component.
	
	This results in the following general stability matrix:
	\begin{align}
	M_1 &= \begin{bmatrix}\displaystyle
	\frac{\partial\mu_1}{\partial N_1} & \cdots & \displaystyle\frac{\partial\mu_1}{\partial N_n} \\
	\vdots & \ddots & \vdots \\
	\displaystyle\frac{\partial\mu_n}{\partial N_1} & \cdots & \displaystyle\frac{\partial\mu_n}{\partial N_n}
	\end{bmatrix} \nonumber\\ 
	&= 
	\begin{bmatrix}
	2B_{11}+\displaystyle\frac{1}{\rho_1} & \cdots & 2B_{1n} \\
	\vdots & \ddots & \vdots \\
	2B_{1n} & \cdots & 2B_{nn} + \displaystyle\frac{1}{\rho_n}
	\end{bmatrix} \label{eq:Firstcriterion}
	\end{align}
	\nomenclature{$M_1$}{Stability matrix}
	When this matrix is positive definite, the mixture is stable \citep{Solokhin2002}. Based on this criterium, when one of the eigenvalues is not positive, the mixture becomes unstable. When the matrix has one zero eigenvalue and is otherwise positive definite, the mixture is on the spinodal and is at the limit of stability \citep{Heidemann1980}.
	
	In case of a binary mixture ($n=2$), the spinodal is also equal to the determinant of matrix $M_1$. For mixtures with more components, this is not always the case anymore, as with an increasing number of components, there are an increasing number of eigenvalues for matrix $M_1$ that can become zero \citep{Solokhin2002}. This can be resolved by checking if the stability matrix is positive definite for small changes of in the concentrations of the components near the concentration where the $det(M_1)$ is zero.
	
	The determinant of matrix $M_1$ for a monodisperse binary mixture is given by:
	\begin{equation}
	det(M_1) = 4(B_{11}B_{22} - B^2_{12})\rho_1\rho_2 + 2B_{11}\rho_1 + 2B_{22}\rho_2 + 1 \label{eq:DetM1}
	\end{equation}
	
	Often however, components are not a 100\% monodisperse. Let's now investigate how the equation for the spinodal of the mixture changes when we introduce polydispersity in one of the components. We define a binary mixture in which one component (component 2) is polydisperse. The concentration of each of the particles in the polydisperse component can be represented by:
	\begin{equation}
	\rho_2 = \begin{bmatrix}x_1 & \cdots & x_{m}
	\end{bmatrix}\times \rho_{2_{tot}} \nonumber
	\end{equation}
	with
	\begin{equation}
	\rho_{2_{tot}} = \sum_i^{m} \rho_{2_i}\nonumber
	\end{equation}
	and
	\begin{equation}
	x = x_1 + \cdots + x_{m} = 1 \nonumber
	\end{equation}
	Each of the components in this mixture has a corresponding virial coefficient and cross virial coefficient.
	\begin{equation}
	B = \begin{bmatrix}
	B_{11} & \cdots & B_{1{m}} \\
	\vdots & \ddots & \vdots \\
	B_{1{m}} & \cdots & B_{mm}
	\end{bmatrix}
	\end{equation}

	Let us investigate the equations for a simple polydisperse mixture, in which the polydisperse component consists out of two sub-components ($a$ and $b$, $n=3$) (figure \ref{fig:GraphicalDiagram}). For the density and for the virial coefficient matrix we obtain:
	\begin{equation}
	\rho_2 = \begin{bmatrix}x_a & x_b
	\end{bmatrix}\times \rho_{2_{tot}} \nonumber
	\end{equation}
	with
	\begin{equation}
	x = x_a + x_b \nonumber
	\end{equation}
	\begin{equation}
	B = \begin{bmatrix}
	B_{11} & B_{12_a} & B_{12_b} \\
	B_{12_a} & B_{2_a2_a} & B_{2_a2_b} \\
	B_{12_b} & B_{2_a2_b} & B_{2_b2_b} \label{eq:Bab}
	\end{bmatrix}
	\end{equation}
	The stability matrix becomes:
	\begin{equation}
	M_1=\left[\begin{matrix}
	2B_{11} + \displaystyle\frac{1}{\rho_1} &  2B_{12_a} & 2B_{12_b}\\[3ex]
	2B_{12_a} & 2B_{2_a2_a}+ \displaystyle\frac{1}{\rho_{2_a}} & 2B_{2_a2_b} \\[3ex]
	2B_{12_b} & 2B_{2_a2_b} & 2B_{2_b2_b}+ \displaystyle\frac{1}{\rho_{2_b}}
	\end{matrix}\right]
	\end{equation}
	This results in the following determinant for the stability matrix:
	\begin{align}
	det(M_1) &= 2B_{11}\rho_1 + 2x_aB_{2_a2_a}\rho_2 + 2x_bB_{2_b2_b}\rho_2 \nonumber\\
	&\quad- 4x_aB_{12_a}^2\rho_1\rho_2 - 4x_bB_{12_b}^2\rho_1\rho_2 \nonumber \\
	&\quad- 4x_ax_bB_{2_a2_b}^2\rho_2^2 + 4x_aB_{11}B_{2_a2_a}\rho_1\rho_2 \nonumber\\
	&\quad+ 4x_bB_{11}B_{2_b2_b}\rho_1\rho_2 + 4x_ax_bB_{2_a2_a}B_{2_b2_b}\rho_2^2\nonumber\\
	&\quad - 8x_ax_bB_{11}B_{2_a2_b}^2\rho_1\rho_2^2 \nonumber\\
	&\quad- 8x_ax_bB_{12_b}^2B_{2_a2_a}\rho_1\rho_2^2 \nonumber\\
	&\quad- 8x_ax_bB_{12_a}^2B_{2_b2_b}\rho_1\rho_2^2 \nonumber\\
	&\quad+ 16x_ax_bB_{12_a}B_{12_b}B_{2_a2_b}\rho_1\rho_2^2 \nonumber\\
	&\quad+ 8x_ax_bB_{11}B_{2_a2_a}B_{2_b2_b}\rho_1\rho_2^2 + 1 \label{eq:DetM1pol}
	\end{align}	
	With increasing number of sub-components, the number of terms in this determinant increases rapidly. This forms an incentive to try and treat the polydisperse component as if it is effectively one component. A natural and convenient choice for this route is coupled to the experimental determination of virial coefficients using membrane osmometry \cite{Ersch2016}. Namely, membrane osmometry yields values that are number averaged. Thus we choose number averaged virial coefficients.
	
	The number averaged virial coefficient of a mixture can be written as:
	\begin{equation}
	\begin{aligned}
	B_{mix} &= B_{11}x_{1}^2 + 2B_{12}x_{1} x_{2} + 2B_{13}x_{1} x_{3}  ... \\
	&= \sum_{i}^{m}\sum_{j}^{m}B_{ij}x_{i}x_{j} \label{eq:Bmix}
	\end{aligned}
	\end{equation}
	in which $B_{ii}$ is the second virial coefficient of the $i$\textsuperscript{th} particle, $B_{ij}$ is the second cross virial coefficient of the $i$\textsuperscript{th} particle and the $j$\textsuperscript{th} particle, and $x_{i}$ is the fraction of the $i$\textsuperscript{th} particle, $\displaystyle\sum x_{i}=1$.
	
	Using this definition, we can map the polydisperse mixture by a $2\times2$ matrix of virial coefficients. We will refer to this  $2\times 2$ matrix of effective virial coefficients.
	\begin{eqnarray}
	B_{11_{eff}} & = & B_{11} \nonumber\\
	B_{12_{eff}} & = & \sum_i^{m}  B[1;1,\cdots,m](i) x_i \nonumber\\
	B_{22_{eff}} & = & \sum_i^{m}\sum_j^{m} x_i B[1,\cdots,m;1,\cdots,m](ij) x_j \nonumber
	\end{eqnarray}
	\begin{equation}
	B_{eff} = \begin{bmatrix}
	B_{11_{eff}} & B_{12_{eff}} \\
	B_{12_{eff}} & B_{22_{eff}} 
	\end{bmatrix}\label{eq:Beff}
	\end{equation}
	For the mixture we considered in eq. \ref{eq:Bab}, the effective virial coefficients become:
	\begin{eqnarray}
	B_{11_{eff}} & = & B_{11} \nonumber\\
	B_{12_{eff}} & = & x_aB_{12_a} + x_bB_{12_b}\nonumber\\
	B_{22_{eff}} & = & x_a^2B_{2_a2_a} + 2x_ax_bB_{2_a2_b}+x_b^2B_{2_b2_b}\label{eq:Beff2}
	\end{eqnarray}
	The effective stability matrix for this mixture becomes then:
	\begin{equation}
	M_{1_{eff}}=\left[\begin{matrix}
	2B_{11_{eff}} + \displaystyle\frac{1}{\rho_1} &  2B_{12_{eff}}\\[3ex]
	2B_{12_{eff}} &  2B_{22_{eff}}+ \displaystyle\frac{1}{\rho_2}
	\end{matrix}\right]
	\end{equation}
	and the determinant then becomes:
	\begin{align}
	det(M_{1_{eff}}) &= 4(B_{11_{eff}}B_{22_{eff}} - B^2_{12_{eff}})\rho_1\rho_2\nonumber\\
	&\quad + 2B_{11_{eff}}\rho_1 + 2B_{22_{eff}}\rho_2 + 1\nonumber\\
	&=4\left(B_{11}(x_a^2B_{2_a2_a} + 2x_ax_bB_{2_a2_b}+x_b^2B_{2_b2_b}) \right.\nonumber\\
	&\quad\quad\left.- (x_aB_{12_a} + x_bB_{12_b})^2\right)\rho_1\rho_2 \nonumber\\
	&\quad+ 2B_{11}\rho_1 + 2\left(x_a^2B_{2_a2_a} + 2x_ax_bB_{2_a2_b}\right.\nonumber\\
	&\quad\quad\left.+x_b^2B_{2_b2_b}\right)\rho_2 + 1\nonumber\\
	&=2B_{11}\rho_1+2x_a^2B_{2_a2_a}\rho_2+2x^2_bB_{2_b2_b}\rho_2\nonumber\\
	&\quad-4x^2_aB_{12_a}^2\rho_1\rho_2-4x^2_bB_{12_b}^2\rho_1\rho_2 \nonumber\\
	&\quad+4x^2_aB_{11}B_{2_a2_a}\rho_1\rho_2+4x^2_bB_{11}B_{2_b2_b}\rho_1\rho_2\nonumber\\
	&\quad+4x_ax_bB_{2_a2_b}\rho_2+8x_ax_bB_{11}B_{2_a2_b}\rho_1\rho_2\nonumber\\
	&\quad-8x_ax_bB_{12_b}B_{12_a}\rho_1\rho_2+1 \label{eq:EffDet}
	\end{align}
	It is clear that eq. \ref{eq:DetM1pol} and eq. \ref{eq:EffDet} are different. Using the effective virial coefficients to determine stability of the mixture possibly results in deviations.
	\subsection{\label{ss:critical}Critical points}
	In a binary mixture, the critical point is a stable point which lies on the stability limit (spinodal) \cite{Heidemann1980} and where the phase boundary and spinodal coincide. In mixtures of more components these become plait points. Critical points and plait points are in general concentrations at which two phases in equilibrium become indistinguishable \cite{Heidemann2013}.
	
	There are two criteria that have to be used to find critical points. The first one is $det(M_1)=0$, which is the equation for the spinodal. The other criterion is based on the fact that at the critical point, the third derivative of the free energy should also be zero. For a multicomponent system, this criterion can be reformulated using Legendre transforms as $det(M_2) = 0$ \cite{Beegle1974}\cite{Reid1977}, where:
	\begin{equation}
	M_2=\begin{bmatrix}
	\displaystyle\frac{\partial\mu_1}{\partial N_1} & \cdots & \displaystyle\frac{\partial\mu_n}{\partial N_n} \\
	\vdots & \ddots & \vdots \\
	\displaystyle\frac{\partial M_1}{\partial N_1} & \cdots & \displaystyle\frac{\partial M_1}{\partial N_n}
	\end{bmatrix} \label{eq:Secondcrit}
	\end{equation}
	Matrix $M_2$ is matrix $M_1$ with one of the rows replaced by the partial derivatives of the determinant of matrix $M_1$. 
	Note: it does not matter which row of the matrix is replaced. 
	
	For a monodisperse binary mixture, this results in the following two matrices for the critical point:
	\begin{equation}
	M_1=\left[\begin{matrix}
	2B_{11} + \displaystyle\frac{1}{\rho_1} &  2B_{12} \\[3ex]
	2B_{12} & 2B_{22}+\displaystyle\frac{1}{\rho_2}
	\end{matrix}\right] \nonumber
	\end{equation}
	and
	\begin{equation}
	M_2=\left[\begin{matrix}
	2B_{11} + \displaystyle\frac{1}{\rho_1} &  2B_{12} \\[3ex]
	\displaystyle-\frac{2B_{22}\rho_2 + 1}{\rho_1^2\rho_2} &\displaystyle-\frac{2B_{11}\rho_1 + 1}{\rho_1\rho_2^2}
	\end{matrix}\right] \nonumber
	\end{equation}
	The set of equations that needs be solved for the critical point is:
	\begin{equation}
	\left\{
	\begin{array}{ll}
	det(M_1) = 1 + 2B_{11}\rho_1 + 2B_{22}\rho_2 - 4B_{12}^2\rho_1\rho_2 \nonumber\\
	\quad\quad\quad\quad\quad\quad\quad\quad\quad+ 4B_{11}B_{22}\rho_1\rho_2 = 0\\
	det(M_2) = (2B_{12}\rho_2)(1+2B_{22}\rho_2)\nonumber\\
	\quad\quad\quad\quad\quad\quad\quad\quad\quad-(1+2B_{11}\rho_1)^2 = 0
	\end{array}
	\right.
	\end{equation}
	For the earlier considered polydisperse mixture containing the two sub-components ($a$ and $b$, $n=3$) we obtain:
	\begin{equation}
	M_2=\left[\begin{matrix}
	2B_{11} + \displaystyle\frac{1}{\rho_1} &  2B_{12_a} & 2B_{12_b}\\[3ex]
	2B_{12_a} & 2B_{2_a2_a}+ \displaystyle\frac{1}{\rho_{2_a}} & 2B_{2_a2_b} \\[3ex]
	P_1 & P_2 & P_3
	\end{matrix}\right]
	\end{equation}
	with 
	\begin{align*}
	P_1 & = -\dfrac{\left(\splitdfrac{- 4x_ax_bB_{2_a2_b}^2\rho_2^2 + 2x_aB_{2_a2_a}\rho_2} { + 2x_bB_{2_b2_b}\rho_2+ 4x_ax_bB_{2_a2_a}B_{2_b2_b}\rho^2_2 + 1}\right)}{x_ax_b\rho_1^2\rho_2^2} \\
	P_2 & = -\dfrac{\left(\splitdfrac{- 4x_bB_{12_b}^2\rho_1\rho_2 + 2B_{11}\rho_1 + 2x_bB_{2_b2_b}\rho_2 }{ + 4x_bB_{11}B_{2_b2_b}\rho_1\rho_2 + 1}\right)}{x^2_ax_b\rho_1\rho_2^3)} \\
	P_3 & = -\dfrac{\left(\splitdfrac{- 4x_aB_{12_a}^2\rho_1\rho_2 + 2B_{11}\rho_1 + 2x_aB_{2_a2_a}\rho_2 }{+ 4x_aB_{11}B_{2_a2_a}\rho_1\rho_2 + 1}\right)}{x_ax_b^2\rho_1\rho_2^3}
	\end{align*}
	In general, $P_i$ can be found using the following equation:
	\begin{equation}
	P_i = - \frac{1}{\rho_i^2}M_{1,(ii)}
	\end{equation}
	in which, $M_{1,(ii)}$ is the minor of matrix $M_1$ at the $i^{th}$-row and $i^{th}$-column.
	
	Combining $det(M_1)$ and $det(M_2)$ results in the following set of equations:
	\begin{equation}
		\left\{
		\begin{array}{ll}
		&det(M_1) = 2B_{11}\rho_1 + 2x_aB_{2_a2_a}\rho_2 + 2x_bB_{2_b2_b}\rho_2 \nonumber\\
		&\quad\quad- 4x_aB_{12_a}^2\rho_1\rho_2 - 4x_bB_{12_b}^2\rho_1\rho_2 \nonumber\\
		&\quad\quad- 4x_ax_bB_{2_a2_b}^2\rho_2^2 + 4x_aB_{11}B_{2_a2_a}\rho_1\rho_2 \nonumber\\
		&\quad\quad+ 4x_bB_{11}B_{2_b2_b}\rho_1\rho_2 + 4x_ax_bB_{2_a2_a}B_{2_b2_b}\rho_2^2 \nonumber\\
		&\quad\quad- 8x_ax_bB_{11}B_{2_a2_b}^2\rho_1\rho_2^2 -8x_ax_bB_{12_b}^2B_{2_a2_a}\rho_1\rho_2^2 \nonumber\\
		&\quad\quad- 8x_ax_bB_{12_a}^2B_{2_b2_b}\rho_1\rho_2^2\nonumber\\
		&\quad\quad + 16x_ax_bB_{12_a}B_{12_b}B_{2_a2_b}\rho_1\rho_2^2 \nonumber\\
		&\quad\quad+ 8x_ax_bB_{11}B_{2_a2_a}B_{2_b2_b}\rho_1\rho_2^2 + 1 = 0\\
		&det(M_2) = 16x_a^2B_{2_a2_a}B_{11}^2B_{2_a2_b}\rho_1^2\rho_2^2 \nonumber\\
		&\quad\quad+ 8x_aB_{11}^2B_{2_a2_b}\rho_1^2\rho_2- 16x_b^2B_{11}^2B_{2_b2_b}^2\rho_1^2\rho_2^2\nonumber\\
		&\quad\quad- 16x_bB_{11}^2B_{2_b2_b}\rho_1^2\rho_2+ 16x_bB_{11}B_{12_b}^2\rho_1^2\rho_2\nonumber\\
		&\quad\quad- 4B_{11}^2\rho_1^2 - 16x_a^2B_{11}B_{12_a}^2B_{2_a2_b}\rho_1^2\rho_2^2\nonumber\\
		&\quad\quad- 16x_a^2B_{2_a2_a}B_{11}B_{12_a}B_{12_b}\rho_1^2\rho_2^2 \nonumber\\
		&\quad\quad- 8x_aB_{11}B_{12_a}B_{12_b}\rho_1^2\rho_2 \nonumber\\
		&\quad\quad+ 32x_b^2B_{11}B_{12_b}^2B_{2_b2_b}\rho_1^2\rho_2^2 \nonumber\\
		&\quad\quad+ 16x_a^2B_{2_a2_a}B_{11}B_{2_a2_b}\rho_1\rho_2^2  \nonumber\\
		&\quad\quad+8x_aB_{11}B_{2_a2_b}\rho_1\rho_2 \nonumber\\
		&\quad\quad- 16x_b^2B_{11}B_{2_b2_b}^2\rho_1\rho_2^2 - 16x_bB_{11}B_{2_b2_b}\rho_1\rho_2 \nonumber\\
		&\quad\quad- 4B_{11}\rho_1 + 16x_a^2B_{12_a}^3B_{12_b}\rho_1^2\rho_2^2 \nonumber\\
		&\quad\quad- 8x_a^2B_{2_a2_a}B_{12_a}B_{12_b}\rho_1\rho_2^2 \nonumber\\
		&\quad\quad- 4x_aB_{12_a}B_{12_b}\rho_1\rho_2 - 8x_a^2B_{12_a}^2B_{2_a2_b}\rho_1\rho_2^2\nonumber\\
		&\quad\quad- 16x_a^2x_b^2B_{12_a}B_{2_a2_b}^2B_{2_b2_b}\rho_2^4 \nonumber\\
		&\quad\quad- 8x_a^2x_bB_{12_a}B_{2_a2_b}^2\rho_2^3 + 8x_ax_b^2B_{12_a}B_{2_b2_b}^2\rho_2^3 \nonumber\\
		&\quad\quad+ 16x_a^2x_b^2B_{2_a2_a}B_{12_a}B_{2_b2_b}^2\rho_2^4  \nonumber\\
		&\quad\quad+ 16x_a^2\rho_bB_{2_a2_a}B_{12_a}B_{2_b2_b}\rho_2^3\nonumber\\
		&\quad\quad+ 8x_ax_bB_{12_a}B_{2_b2_b}\rho_2^2 + 4x_a^2B_{2_a2_a}B_{12_a}\rho_2^2 \nonumber\\
		&\quad\quad+ 2x_aB_{12_a}\rho_2 - 16x_b^2B_{12_b}^4\rho_1^2\rho_2 \nonumber\\
		&\quad\quad+ 16x_b^2B_{12_b}^2B_{2_b2_b}\rho_1\rho_2^2  + 8x_bB_{12_b}^2\rho_1\rho_2 \nonumber\\
		&\quad\quad+ 16x_a^2x_b^2B_{12_b}B_{2_a2_b}^3\rho_2^4 \nonumber\\
		&\quad\quad- 16x_a^2x_b^2B_{2_a2_a}B_{12_b}B_{2_a2_b}B_{2_b2_b}\rho_2^4 \nonumber\\
		&\quad\quad- 8x_ax_b^2B_{12_b}B_{2_a2_b}B_{2_b2_b}\rho_2^3 \nonumber\\
		&\quad\quad- 8x_a^2x_bB_{2_a2_a}B_{12_b}B_{2_a2_b}\rho_2^3\nonumber\\
		&\quad\quad- 4x_ax_bB_{12_b}B_{2_a2_b}\rho_2^2 + 4x_a^2B_{2_a2_a}B_{2_a2_b}\rho_2^2 \nonumber\\
		&\quad\quad+ 2x_aB_{2_a2_b}\rho_2 - 4x_b^2B_{2_b2_b}^2\rho_2^2 \nonumber\\
		&\quad\quad- 4x_bB_{2_b2_b}\rho_2 - 1 = 0
		\end{array}
		\right.
	\end{equation}
	Since in this set of equations there are more higher order terms present, it is possible that this results in multiple plait points, depending on the concentration of each of the components in the mixture. Care should be taken that the solutions of the set of equations are concentrations are at the limit of stability, this can be done by checking the eigenvalues of the stability matrix.
	
	If we use the effective virial coefficients for this mixture as defined in eq. \ref{eq:Beff2}, we obtain:
	\begin{equation}
	M_{2_{eff}}=\left[\begin{matrix}
	2B_{11_{eff}} + \displaystyle\frac{1}{\rho_1} &  2B_{12_{eff}}\\[3ex]
	-\displaystyle\frac{2B_{22_eff}\rho_2 + 1}{\rho_1^2\rho_2} &  -\displaystyle\frac{2B_{11_eff}\rho_1+1}{\rho_1\rho_2^2}
	\end{matrix}\right] \nonumber
	\end{equation}
	The determinant of this matrix becomes:
	\begin{align}
	det(M_{2_{eff}}) &= - 4B_{11_eff}^2\rho_1^2 - 4B_{11_eff}\rho_1\nonumber\\
	&\quad + 4B_{12_eff}B_{22_eff}\rho_2^2 + 2B_{12_eff}\rho_2 - 1\nonumber\\
	&= - 4B_{11}^2\rho_1^2 - 4B_{11}\rho_1 + 4(x_aB_{12_a}+x_bB_{12_b})\nonumber\\
	&\quad\times(x^2_aB_{2_a2_b}+2x_ax_bB_{2_a2_b}\nonumber\\
	&\quad+x^2_bB_{2_b2_b})\rho_2^2 + \nonumber\\
	&\quad 2(x_aB_{12_a}+x_bB{12_b})\rho_2 - 1\nonumber\\
	&=-4B_{11}^2\rho^2_1-4B_{11}\rho_1+4x_a^3B_{12_a}B_{2_a2_a}\rho_2^2\nonumber\\
	&\quad+8x^2_ax_bB_{12_a}B_{2_a2_b}\rho^2_2+4x_ax_b^2B_{12_a}B_{2_b2_b}\rho^2_2\nonumber\\
	&\quad +4x^2_ax_bB_{12_b}B_{2_a2_a}\rho_2^2+8 x_ax_b^2B_{12_b}B_{2_a2_b}\rho^2_2\nonumber\\
	&\quad+4x_b^3B_{12_b}B_{2_b2_b}\rho^2_2+2x_aB_{12_a}\rho_2\nonumber\\
	&\quad+ 2x_bB_{12_b}\rho_2-1 \nonumber
	\end{align}
	Which results in the following system of equations for the critical point:
	\begin{equation}
	\left\{
	\begin{array}{ll}
	&det(M_{1_eff}) = 2B_{11}\rho_1+2x_a^2B_{2_a2_a}\rho_2+ 2x^2_bB_{2_b2_b}\rho_2\nonumber\\
	&\quad\quad-4x^2_aB_{12_a}^2\rho_1\rho_2 -4x^2_bB_{12_b}^2\rho_1\rho_2\nonumber\\
	&\quad\quad+ 4x^2_aB_{11}B_{2_a2_a}\rho_1\rho_2 + 4x^2_bB_{11}B_{2_b2_b}\rho_1\rho_2\nonumber\\
	&\quad\quad+4x_ax_bB_{2_a2_b}\rho_2+8x_ax_bB_{11}B_{2_a2_b}\rho_1\rho_2\nonumber\\
	&\quad\quad-8x_ax_bB_{12_b}B_{12_a}\rho_1\rho_2+1 = 0\\
	&det(M_{2_eff}) =-4B_{11}^2\rho^2_1-4B_{11}\rho_1\nonumber\\
	&\quad\quad+4x_a^3B_{12_a}B_{2_a2_a}\rho_2^2+8x^2_ax_bB_{12_a}B_{2_a2_b}\rho^2_2\nonumber\\
	&\quad\quad+4x_ax_b^2B_{12_a}B_{2_b2_b}\rho^2_2+4x^2_ax_bB_{12_b}B_{2_a2_a}\rho_2^2\nonumber\\
	&\quad\quad+8 x_ax_b^2B_{12_b}B_{2_a2_b}\rho^2_2+4x_b^3B_{12_b}B_{2_b2_b}\rho^2_2\nonumber\\
	&\quad\quad+2x_aB_{12_a}\rho_2+ 2x_bB_{12_b}\rho_2-1 = 0
	\end{array}
	\right.
	\end{equation}
	Also for the third derivative of the Helmholtz free energy, we see that reducing the polydispersity by using the effective virial coefficients, results in fewer terms in the equation and possible deviations in determining the critical point.
	\subsection{\label{ss:phaseboundary}Phase boundary}
	When a mixture becomes unstable and phase separates into two or more phases, the chemical potential of each component and the osmotic pressure is the same in all phases \cite{Hill1986}.
	\begin{equation}
	\left\{
	\begin{aligned}
	\beta\Pi^{I} &= \beta\Pi^{II}  &=\cdots\\
	\beta\mu_{1}^{I} &= \beta\mu_{1}^{II} &=\cdots\\
	&\vdots&\\
	\beta\mu_{n}^{I} &= \beta\mu_{n}^{II} &=\cdots\\ 
	\end{aligned}
	\right.
	\end{equation}
	where the phases are denoted by $I,II,...$.
	
	For a system that separates into two phases, we obtain, using eq. \ref{eq:OsmoticGeneral} and eq. \ref{eq:Chempotential}, the following set of equations for the general case of $n$ components:
	\begin{equation}
		\left\{
		\begin{aligned}
		\rho_1^{I} + \cdots + \rho_n^{I} + &\sum_{i}^{n}\sum_{j}^{n}B_{ij}\rho_{i}^{I}\rho_{j}^{I}  \\
		&= \\
		 \rho_1^{II} + \cdots + \rho_n^{II} &+ \sum_{i}^{n}\sum_{j}^{n}B_{ij}\rho_{i}^{II}\rho_{j}^{II} \\
		\ln(\rho_1^{I})+2\left(\sum_j^{n} B_{1j}\rho_j^{I}\right) & =  \ln(\rho_1^{II})+2\left(\sum_j^{n} B_{1j}\rho_j^{II}\right) \\
		&\vdots \\
		\ln(\rho_n^{I})+2\left(\sum_j^{n} B_{nj}\rho_j^{I}\right) & =  \ln(\rho_n^{II})+2\left(\sum_j^{n} B_{nj}\rho_j^{II}\right)
		\end{aligned}
		\right. \nonumber
	\end{equation}
	This set of equations has $2 \times n$ unknowns and $n + 1$ equations. The set of equations can be solved by fixing one of the concentrations for one phase and the ratio of the concentrations of the other components for the same phase. To solve this set of equations in order to find the concentration of each component in each phase, without fixing any of the concentrations, we need therefore an extra set of equations. 
	
	This extra set of equations stems from the fact that during phase separation no particles are lost and no new particles are created. The total number of components in the system is therefore given by:
	\begin{equation}
	N = \sum_i^n N_i^{I}+\sum_i^n N_i^{II} \label{eq:N}
	\end{equation} 
	Also the total volume, $V$, of the system does not change. With the total volume of the system given by:
	\begin{equation}
	V = V^{I} + V^{II} \label{eq:V}
	\end{equation}
	the concentrations of each component in each phase are thus given by:
	\begin{equation}
	\begin{aligned}
	\rho^{I}_1 = \frac{N^{I}_1}{V^{I}}& & \cdots & &\rho^{I}_n = \frac{N^{I}_n}{V^{I}} \\
	\rho^{II}_1 = \frac{N^{II}_1}{V^{II}}& & \cdots & &\rho^{II}_n = \frac{N^{II}_n}{V^{II}} 
	\end{aligned} \label{eq:dens}
	\end{equation}
	The total number of compounds in the system can be found using:
	\begin{equation}
	\rho = \sum_i^n \rho_i = \frac{\sum\limits_i^n N_i}{V} = \frac{\sum\limits_i^n N_i^{I}+\sum\limits_i^n N_i^{II}}{V^{I}+V^{II}} \nonumber
	\end{equation}
	which can be rewritten to:
	\begin{equation}
	\begin{aligned}
	\rho &= \sum_i^n \frac{N_i^I}{V^I+V^{II}} + \sum_i^n \frac{N_i^{II}}{V^I+V^{II}} \\
	&= \frac{V^{I}}{V^{I}+V^{II}} \sum_i^n \frac{N_i^I}{V^I} + \frac{V^{II}}{V^{I}+V^{II}} \sum_i^n \frac{N_i^{II}}{V^I+V^{II}} \\
	&=\alpha \sum_i^{n}\rho_i^{I} + (1-\alpha) \sum_i^{n}\rho_i^{II} = \sum_i^{n}\rho_i \nonumber
	\end{aligned}
	\end{equation}
	with 
	\begin{equation}
	\alpha = \frac{V^{I}}{V^{I}+V^{II}} \label{eq:alpha}
	\end{equation}	
	\nomenclature{$\alpha$}{Volume fraction phase $I$  $\displaystyle\alpha = \frac{V^{I}}{V^{I}+V^{II}}$}
	This results in an extra set of $n$ equations and one more unknown ($\alpha$). The complete set of equations to solve for the binodal then becomes:
	\begin{equation}
		\left\{
		\begin{aligned}
		\rho_1^{I} + \cdots + \rho_n^{I} + &\sum_{i}^{n}\sum_{j}^{n}B_{ij}\rho_{i}^{I}\rho_{j}^{I}  \\
		&= \\
		\rho_1^{II} + \cdots + \rho_n^{II} &+ \sum_{i}^{n}\sum_{j}^{n}B_{ij}\rho_{i}^{II}\rho_{j}^{II} \\
		\ln(\rho_1^{I})+2\left(\sum_j^{n} B_{1j}\rho_j^{I}\right) & =  \ln(\rho_1^{II})+2\left(\sum_j^{n} B_{1j}\rho_j^{II}\right) \\
		&\vdots \\
		\ln(\rho_n^{I})+2\left(\sum_j^{n} B_{nj}\rho_j^{I}\right) & =  \ln(\rho_n^{II})+2\left(\sum_j^{n} B_{nj}\rho_j^{II}\right)\\
		\rho_1 &= \alpha \rho_1^{I} + (1-\alpha) \rho_1^{II} \\
		&\vdots \\
		\rho_n &= \alpha \rho_n^{I} + (1-\alpha) \rho_n^{II}
		\end{aligned}
		\right. \nonumber
	\end{equation}
	Systems for which there are more than two distinguishable components ($n > 2$) can theoretically have more than two coexisting phases, according to the Gibbs phase rule. With an increasing number of phases, the set of equations to solve increases as well. The number of equations needed to solve for an arbitrary number of $f$ phases is: $f \times n + f - 1$. The set of equations then becomes:
	\begin{widetext}
		\begin{equation}
		\left\{
		\begin{aligned}
		\rho_1^{I} + \cdots + \rho_n^{I} + \sum_{i}^{n}\sum_{j}^{n}B_{ij}\rho_{i}^{I}\rho_{j}^{I}  &=  \rho_1^{II} + \cdots + \rho_n^{II} + \sum_{i}^{n}\sum_{j}^{n}B_{ij}\rho_{i}^{II}\rho_{j}^{II} \\
		&\vdots\\
		\rho_1^{f-1} + \cdots + \rho_n^{f-1} + \sum_{i}^{n}\sum_{j}^{n}B_{ij}\rho_{i}^{f-1}\rho_{j}^{f-1}  &=  \rho_1^{f} + \cdots + \rho_n^{f} + \sum_{i}^{n}\sum_{j}^{n}B_{ij}\rho_{i}^{f}\rho_{j}^{f} \\
		\ln(\rho_1^{I})+2\left(\sum_j^{n} B_{1j}\rho_j^{I}\right) & =  \ln(\rho_1^{II})+2\left(\sum_j^{n} B_{1j}\rho_j^{II}\right) \\
		&\vdots \\
		\ln(\rho_1^{f-1})+2\left(\sum_j^{n} B_{1j}\rho_j^{f-1}\right) & =  \ln(\rho_1^{f})+2\left(\sum_j^{n} B_{1j}\rho_j^{f}\right) \\
		&\vdots \\
		\ln(\rho_n^{I})+2\left(\sum_j^{n} B_{nj}\rho_j^{I}\right) & =  \ln(\rho_n^{II})+2\left(\sum_j^{n} B_{nj}\rho_j^{II}\right)\\
		&\vdots\\
		\ln(\rho_n^{f-1})+2\left(\sum_j^{n} B_{nj}\rho_j^{f-1}\right) & =  \ln(\rho_n^{f})+2\left(\sum_j^{n} B_{nj}\rho_j^{f}\right)\\
		\rho_1 = \alpha_1 \rho_1^{I} + \alpha_2 \rho_1^{II} +&\cdots + \left(1-\sum_i^{f-1}\alpha_i\right)\rho_1^f \\
		&\vdots \\
		\rho_n = \alpha_1 \rho_n^{I} + \alpha_2 \rho_n^{II} +&\cdots + \left(1-\sum_i^{f-1}\alpha_i\right)\rho_n^f
		\end{aligned}
		\right. \nonumber
		\end{equation}
		
		with
		\begin{equation}
		\alpha_1 = \frac{V^{I}}{\sum\limits_i^{f}V^{i}} \qquad \cdots \qquad \alpha_{f-1} = \frac{V^{f-1}}{\sum\limits_i^{f}V^{i}} \nonumber
		\end{equation}
		\nomenclature{$f$}{Number of coexisting phases}
		Now that we have the general equations for the phaseboundary, let us investigate the set of equations we need to solve for the mixtures we defined earlier. For a monodisperse binary mixture the set of equations to solve for the phase boundary is given by:
		\begin{equation}
		\left\{
		\begin{aligned}
		\rho_1^{I} + \rho_2^{I} + B_{11}\rho_1^{I^2} + 2B_{12}\rho_1^{I}\rho_2^{I} + B_{22}\rho_2^{I^2} 
		&= 	\rho_1^{II} + \rho_2^{II} + B_{11}\rho_1^{II^2}  +  2B_{12}\rho_1^{II}\rho_2^{II} + B_{22}\rho_2^{II^2}	\\
		\ln(\rho_1^{I})+2B_{11}\rho_1^{I} + 2B_{12}\rho_2^{I} & = \ln(\rho_1^{II})+2B_{11}\rho_1^{II} + 2B_{12}\rho_2^{II}\\
		\ln(\rho_2^{I})+2B_{12}\rho_1^{I} + 2B_{22}\rho_2^{I} & = \ln(\rho_2^{II})+2B_{12}\rho_1^{II} + 2B_{22}\rho_2^{II}\\
		\rho_1 &= \alpha \rho_1^{I} + (1-\alpha) \rho_1^{II} \\
		\rho_2 &= \alpha \rho_2^{I} + (1-\alpha) \rho_2^{II}
		\end{aligned}
		\right. \nonumber
		\end{equation}
		For the polydisperse binary mixture we considered earlier (with two sub-components $a$ and $b$, $n=3$) this set of equations becomes:
		
		\begin{equation}
		\left\{
		\begin{aligned}
		\rho_1^{I} + \rho_{2_a}^{I} + \rho_{2_b}^{I} + B_{11}\rho_1^{I^2} + 2B_{12_a}\rho_1^{I}\rho_{2_a}^{I} + 2B_{12_b}\rho_1^{I}\rho_{2_b}^{I} &+ B_{2_a2_a}\rho_{2_a}^{I^2} + 2B_{2_a2_b}\rho_{2_a}^{I}\rho_{2_a}^{I} + B_{2_b2_b}\rho_{2_b}^{I^2}\\
		=\rho_1^{II} + \rho_{2_a}^{II} + \rho_{2_b}^{II} + B_{11}\rho_1^{II^2} + 2 B_{12_a}\rho_1^{II}\rho_{2_a}^{II} + &2B_{12_b}\rho_1^{II}\rho_{2_b}^{II} +  B_{2_a2_a}\rho_{2_a}^{II^2} + 2 B_{2_a2_b}\rho_{2_a}^{II}\rho_{2_b}^{II} +  B_{2_b2_b}\rho_{2_b}^{II^2}\\
		\ln(\rho_1^{I})+2B_{11}\rho_1^{I} + 2B_{12_a}\rho_{2_a}^{I} + 2B_{12_b}\rho_{2_b}^{I} & = \ln(\rho_1^{II})+2B_{11}\rho_1^{II} + 2B_{12_a}\rho_{2_a}^{II} + 2B_{12_b}\rho_{2_b}^{II} \\
		\ln(\rho_{2_a}^{I})+2B_{12_a}\rho_1^{I} + 2B_{2_a2_a}\rho_{2_a}^{I} + 2B_{2_a2_b}\rho_{2_b}^{I}& = \ln(\rho_{2_a}^{II})+2B_{12_a}\rho_1^{II} + 2B_{2_a2_a}\rho_{2_a}^{II} + 2B_{2_a2_b}\rho_{2_b}^{II}\\
		\ln(\rho_{2_b}^{I})+2B_{12_b}\rho_1^{I} + 2B_{2_a2_b}\rho_{2_a}^{I} + 2B_{2_b2_b}\rho_{2_b}^{I}& = \ln(\rho_{2_b}^{II})+2B_{12_b}\rho_1^{II} + 2B_{2_a2_b}\rho_{2_a}^{II} + 2B_{2_b2_b}\rho_{2_b}^{II}\\
		\rho_1 &= \alpha \rho_1^{I} + (1-\alpha) \rho_1^{II} \\
		\rho_{2_a} &= \alpha \rho_{2_a}^{I} + (1-\alpha) \rho_{2_a}^{II} \\
		\rho_{2_b} &= \alpha \rho_{2_b}^{I} + (1-\alpha) \rho_{2_b}^{II} \nonumber
		\end{aligned}
		\right.
		\end{equation}
		Note that the ratio between $\rho_{2_a}^{I}$ and $\rho_{2_b}^{I}$ is not necessarily the same as the ratio between  $\rho_{2_a}^{II}$ and $\rho_{2_b}^{II}$, since fractionation between the components can occur.
		
		Using the effective virial coefficients we obtain:
		\begin{equation}
		\left\{
		\begin{aligned}
		\rho_1^{I} + \rho_2^{I} + B_{11_{eff}}\rho_1^{I^2} + 2B_{12_{eff}}\rho_1^{I}\rho_2^{I} + B_{22_{eff}}\rho_2^{I^2} &= 	\rho_1^{II} + \rho_2^{II} + B_{11_{eff}}\rho_1^{II^2} + 2B_{12_{eff}}\rho_1^{II}\rho_2^{II} + B_{22_{eff}}\rho_2^{II^2}	\\
		\ln(\rho_1^{I})+2B_{11_{eff}}\rho_1^{I} + 2B_{12_{eff}}\rho_2^{I} & = \ln(\rho_1^{II})+2B_{11_{eff}}\rho_1^{II} + 2B_{12_{eff}}\rho_2^{II}\\
		\ln(\rho_2^{I})+2B_{12_{eff}}\rho_1^{I} + 2B_{22_{eff}}\rho_2^{I} & = \ln(\rho_2^{II})+2B_{12_{eff}}\rho_1^{II} + 2B_{22_{eff}}\rho_2^{II}\\
		\rho_1 &= \alpha \rho_1^{I} + (1-\alpha) \rho_1^{II} \\
		\rho_2 &= \alpha \rho_2^{I} + (1-\alpha) \rho_2^{II}
		\end{aligned}
		\right.\nonumber
		\end{equation}
	\end{widetext}
	It is clear from the equations, that when using the effective virial coefficients for calculating the phase boundary all information about the (changes in) distribution of the polydisperse component becomes untraceable.
	
	\section{\label{s:RD}Results and discussion}
	
\begin{figure}
		\centering
		\includegraphics[trim=1cm 0cm 1cm 0cm, clip=true ,width=0.42\textwidth]{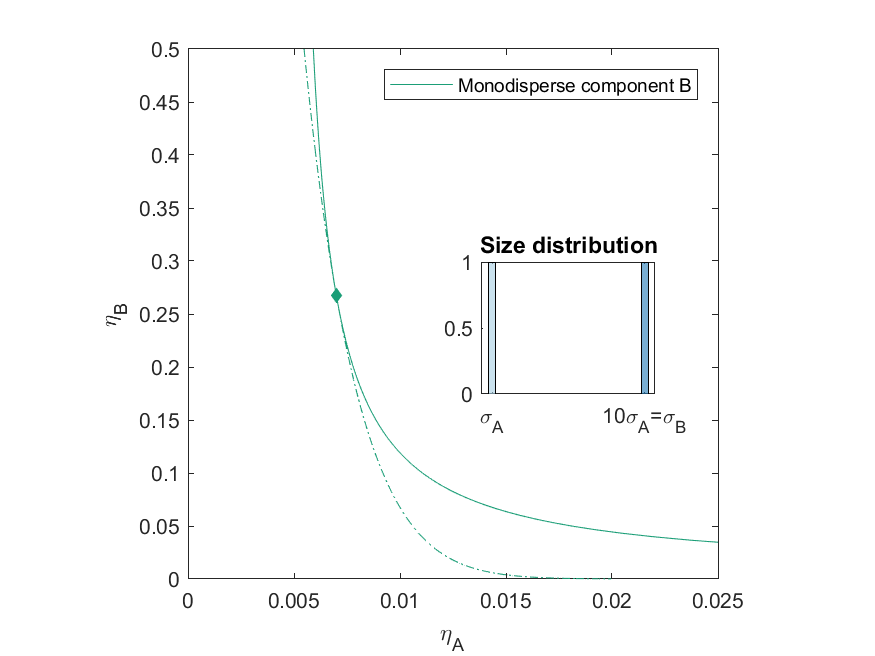}
		\caption{Phase diagram for monodisperse binary (component $A$ and $B$) additive hard sphere mixture with size ratio $q = \sigma_A / \sigma_B =  1/10$, plotted as a function of the partial packing fractions, $\eta_A$ and $\eta_B$. The spinodal (solid line) and binodal (dashed line) meet each other at the critical point (diamond)}
		\label{fig:monodisperse}
	\end{figure}

	In this work we calculated the phase diagram for a variety of binary additive mixtures of a small hard sphere $A$ and a larger hard sphere $B$ with a size ratio $q = \sigma_A / \sigma_B =  1/10$. We started by calculating the phase diagram of this monodisperse mixture (figure \ref{fig:monodisperse}) and gradually introduced polydispersity into the composition of component $B$ (figure \ref{fig:PD5050smal} trough \ref{fig:9bins}). Component $B$ is characterized by a degree in polydispersity ($PD$), defined by:
	\begin{equation}
	PD = \frac{\sqrt{\sum{(\sigma_{B_i} - \sigma_B)^2}\times N_{B_i}/N_B}}{\sigma_B}\times 100\nonumber 
	\end{equation}
	\nomenclature{$q$}{Size ratio between components $\displaystyle\frac{\sigma_A}{\sigma_B}$}
	\nomenclature{$\eta$}{Packing fraction $\displaystyle\frac{\pi \rho \sigma^3}{6}$}
	\nomenclature{$PD$}{Degree in polydispersity}
	For all particles, the concentrations are expressed as a dimensionless parameter according to $\displaystyle \eta = \frac{\pi \rho \sigma^3}{6}$. We calculated the critical point, the phase separation boundary, and the spinodal of the various mixtures. Next to that, we also investigated the composition of the child phases, volume ratio between the phases ($\alpha$), and the fractionation of the polydisperse component $B$ for a specific parent mixture. 
	
	In order to be able to compare our results with the monodisperse case we first refer to figure \ref{fig:monodisperse}, where we see that the binary mixture phase separates at very low volume fractions of component $A$ ($\eta_{A_{crit}} = 0.007$) and significantly higher concentrations of component $B$ ($\eta_{B_{crit}} = 0.267$). 
	
	\subsection{\label{ss:2pseudo}Polydisperse mixtures with 2 sub-components}
	
\begin{figure}
		\centering
		\includegraphics[trim=1cm 0cm 1cm 0cm, clip=true ,width=0.42\textwidth]{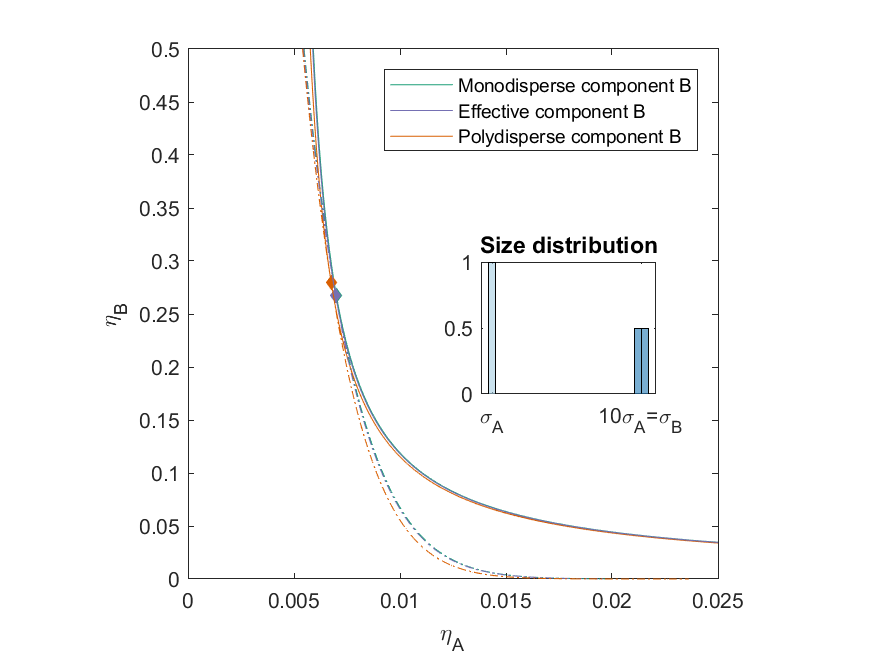}
		\caption{Phase diagram for binary (component $A$ and $B$) additive hard sphere mixture with size ratio $q = \sigma_A / \sigma_B =  1/10$, component $A$ is monodisperse, component $B$ is polydisperse (see size distribution), with a number average size 10 times the size of component $A$, plotted as a function of the partial packing fractions, $\eta_A$ and $\eta_B$. The spinodal (solid line) and binodal (dashed line) meet each other at the critical point (diamond)}
		\label{fig:PD5050smal}
	\end{figure}

	In figure \ref{fig:PD5050smal} we show the phase diagram for the case of a slight polydispersity in component $B$. Component $B$ consists of two sub-components and has a $PD = 4.00$. These components are additive spheres in two sizes (both present in the same amount), with the number average size of the mixture equal to that of the size of the monodisperse mixture of figure \ref{fig:monodisperse}. The mixture therefore consist of three components. We calculated the phase diagram using both the simplified $2 \times 2$ effective virial coefficient matrix described in the theory (we refer to this as the effective mixture) and the full $3 \times 3$ virial coefficient matrix (to which we refer as the polydisperse mixture). The difference between the phase boundary, spinodal and critical point of the monodisperse mixture and the effective mixture is negligible. We see however that the introduction of the polydispersity causes the critical point to shift to a higher volume fraction of component $B$ ($\eta_{A_{crit}} =0.007, \eta_{B_{crit}} =0.280$) and that especially at lower volume fraction of component $B$ the phase separation boundary shifts towards slightly lower packing fractions.
	
	\begin{figure}
		\centering
		\includegraphics[trim=1cm 0cm 1cm 0cm, clip=true,width=0.42\textwidth]{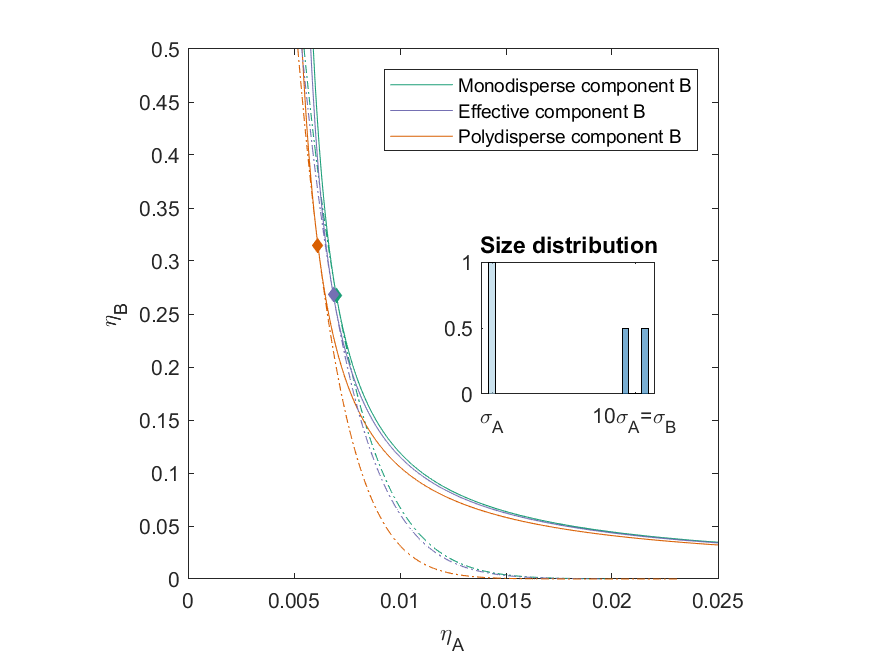}
		\caption{ Phase diagram for binary (component $A$ and $B$) additive hard sphere mixture with size ratio $q = \sigma_A / \sigma_B =  1/10$, component $A$ is monodisperse, component $B$ is polydisperse (see size distribution, SD twice of mixture in figure \ref{fig:PD5050smal}), with a number average size 10 times the size of component $A$, plotted as a function of the partial packing fractions, $\eta_A$ and $\eta_B$. The spinodal (solid line) and binodal (dashed line) meet each other at the critical point (diamond)}
		\label{fig:PD5050big}
	\end{figure}
	
	\begin{figure*}
		\centering
		\subfloat[Large amount of larger spheres\label{fig:PD2575}]{%
			\includegraphics[trim=1cm 0cm 1cm 0cm, clip=true,width=0.42\textwidth]{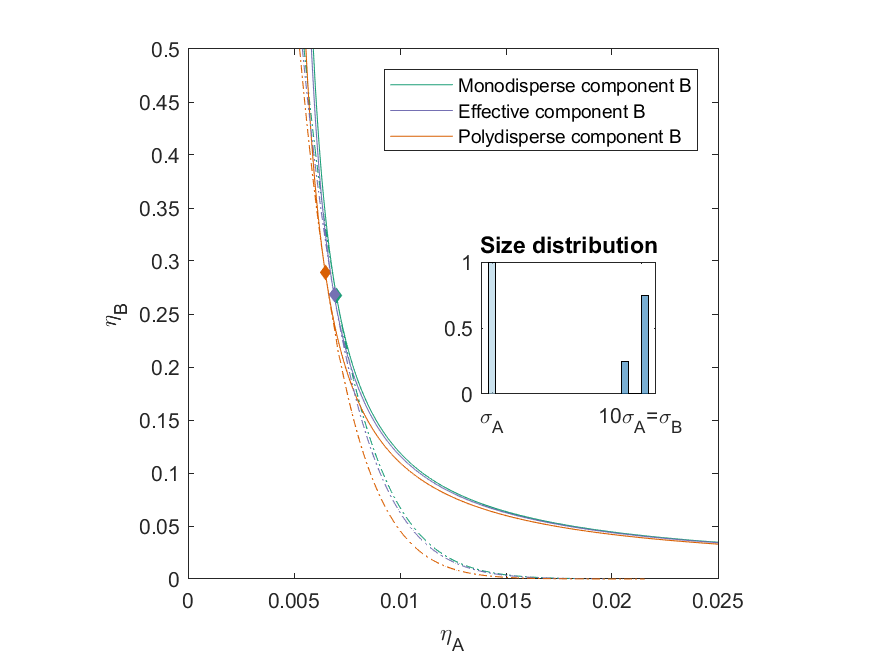}%
		}
		\subfloat[Small amount of larger spheres\label{fig:PD7525}]{%
			\includegraphics[trim=1cm 0cm 1cm 0cm, clip=true,width=0.42\textwidth]{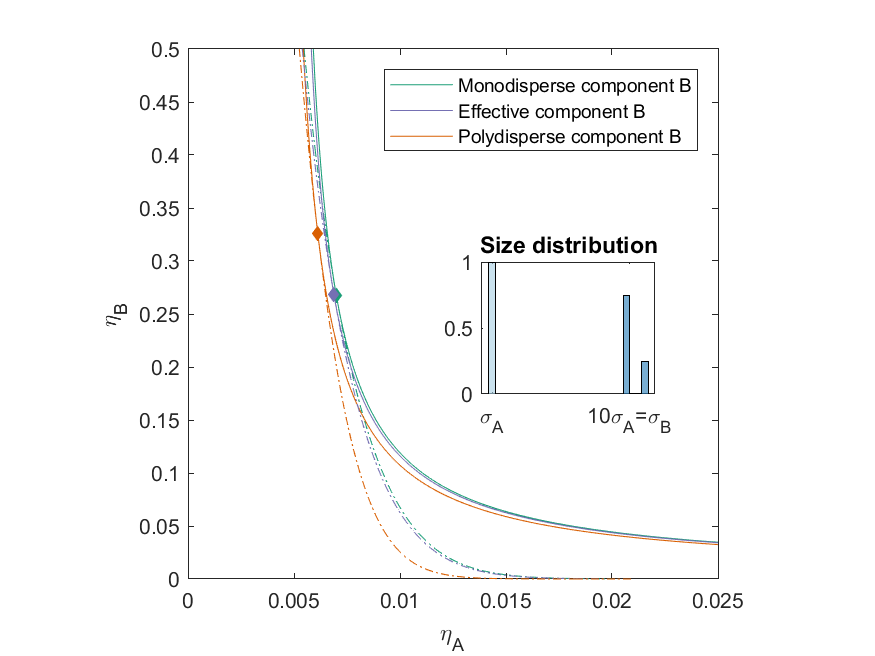}%
		}\hfill
		\subfloat[Large amount of larger spheres\label{fig:PD1090}]{%
			\includegraphics[trim=1cm 0cm 1cm 0cm, clip=true,width=0.42\textwidth]{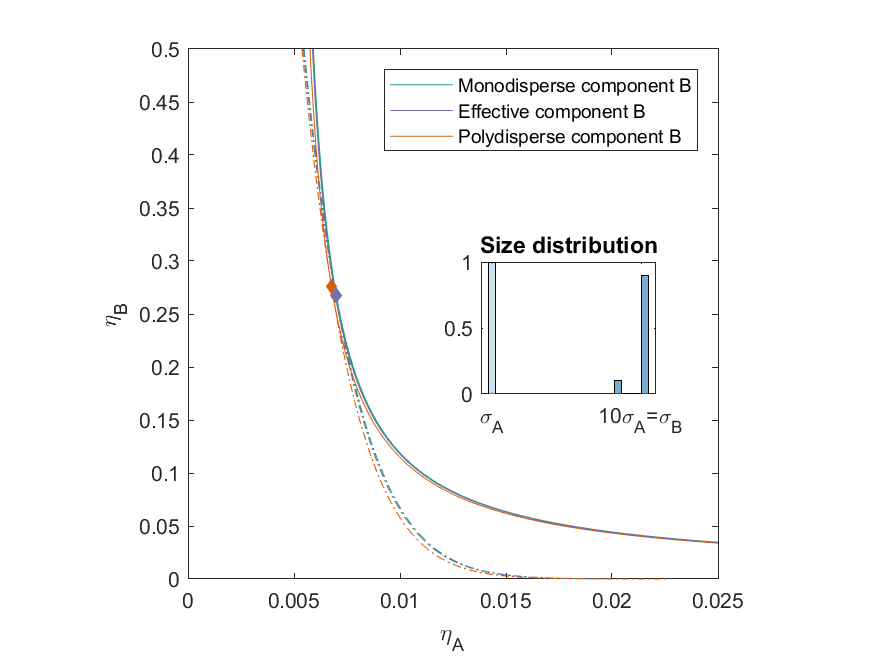}%
		}
		\subfloat[Small amount of larger spheres\label{fig:PD9010}]{%
			\includegraphics[trim=1cm 0cm 1cm 0cm, clip=true,width=0.42\textwidth]{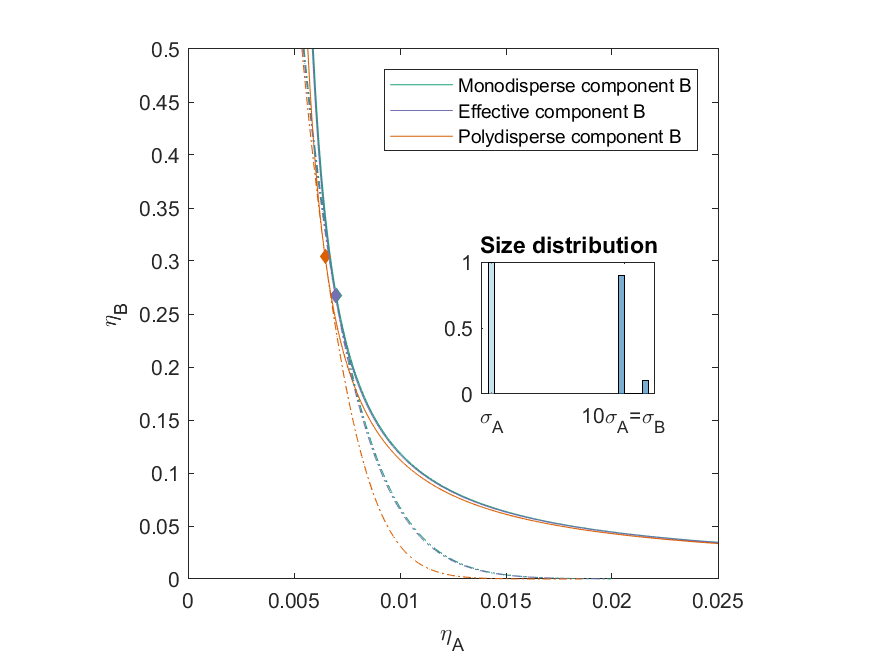}%
		}
		\caption{Phase diagram for binary (component $A$ and $B$) additive hard sphere mixture with size ratio $q = \sigma_A / \sigma_B =  1/10$, component $A$ is monodisperse, component $B$ is polydisperse (see size distribution), with a number average size 10 times the size of component $A$, plotted as a function of the partial packing fractions, $\eta_A$ and $\eta_B$. The spinodal (solid line) and binodal (dashed line) meet each other at the critical point (diamond)}
		\label{fig:PDskew}
	\end{figure*}

	For figure \ref{fig:PD5050big} we increased the size difference between the smaller and the larger spheres for component $B$, with the standard deviation twice the standard deviation of the spheres in figure \ref{fig:PD5050smal} and the $PD = 8.00$. The patterns we saw in figure \ref{fig:PD5050smal} are more pronounced for this mixture: the increase in size difference causes the critical point to shift to higher packing fractions for component $B$ and lower packing fractions for component $A$ ($\eta_{A_{crit}} =0.006, \eta_{B_{crit}} =0.315$). The phase boundary, i.e. the binodal, shifts to significantly lower packing fractions, especially at lower volume concentrations of component $B$. For this mixture we see also that the spinodal of the polydisperse mixture shifts towards lower packing fractions of $A$. There is a slight difference between the positions of the binodal, spinodal and critical point of the monodisperse mixture and effective mixture.

	In figure \ref{fig:PDskew} we introduced skewness in the size distribution of component $B$. For the first two mixtures (\ref{fig:PD2575} and \ref{fig:PD7525}) the ratio between the bigger and the smaller sub-component was $25/75$. The polydispersity for both mixtures is the same, $PD = 6.93$. For the other two mixtures (\ref{fig:PD1090} and \ref{fig:PD9010}), the ratio between the bigger and the smaller sub-component was more extreme, namely $90/10$, with $PD = 4.80$. For all these mixtures, we find that the critical point shifts towards higher packing fraction of component $B$ (($\eta_{A_{crit}} =0.006, \eta_{B_{crit}} 0.290$) for figure \ref{fig:PD2575}, ($\eta_{A_{crit}} =0.006, \eta_{B_{crit}} 0.326$) for figure \ref{fig:PD7525}, ($\eta_{A_{crit}} =0.007, \eta_{B_{crit}} 0.276$) for figure \ref{fig:PD1090}, and ($\eta_{A_{crit}} =0.007, \eta_{B_{crit}} 0.304$) for figure \ref{fig:PD9010}).

	Next to this shift in critical point, we also see that the spinodal shifts towards lower packing fractions of component $A$, and that the binodal shifts towards lower packing fraction for lower concentrations of component $B$, although there is a difference in the amount of shift. We kept the number average size of the spheres for component $B$ the same for all mixtures, meaning that the largest spheres in the mixture in figures \ref{fig:PD7525} and \ref{fig:PD9010} are larger than the largest spheres in the mixture in figures \ref{fig:PD2575} and \ref{fig:PD1090}. From this we can conclude that the larger spheres, even though they are smaller in number, have a higher impact on the concentration of the critical point and the position of the phase boundary.
	
	Looking at the different mixtures (figures \ref{fig:PD5050smal} through \ref{fig:PDskew}) we can conclude that polydispersity shifts the critical point to higher packing fraction for component $B$ and lower packing fractions of component $A$ compared to a monodisperse mixture with the same average sizes. The phase boundary shifts to lower packing fractions at lower concentrations of component $B$. The shift is dependent on the size distribution and then in a large part on the size and the concentration of the largest particle in the mixture.
	
	\subsection{\label{ss:9pseudo}Polydispersity with 9 sub-components}
	
\begin{figure*}
		\centering
		\subfloat[Narrow Gaussian distribution\label{fig:smallgaus}]{%
			\includegraphics[trim=1cm 0cm 1cm 0cm, clip=true,width=0.42\textwidth]{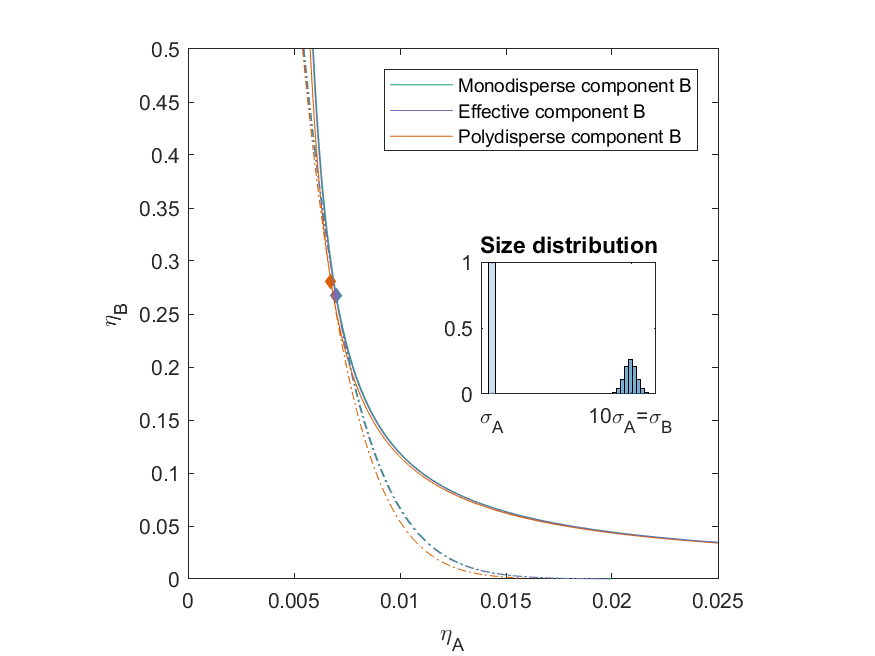}%
		}
		\subfloat[Broad Gaussian distribution\label{fig:biggaus}]{%
			\includegraphics[trim=1cm 0cm 1cm 0cm, clip=true,width=0.42\textwidth]{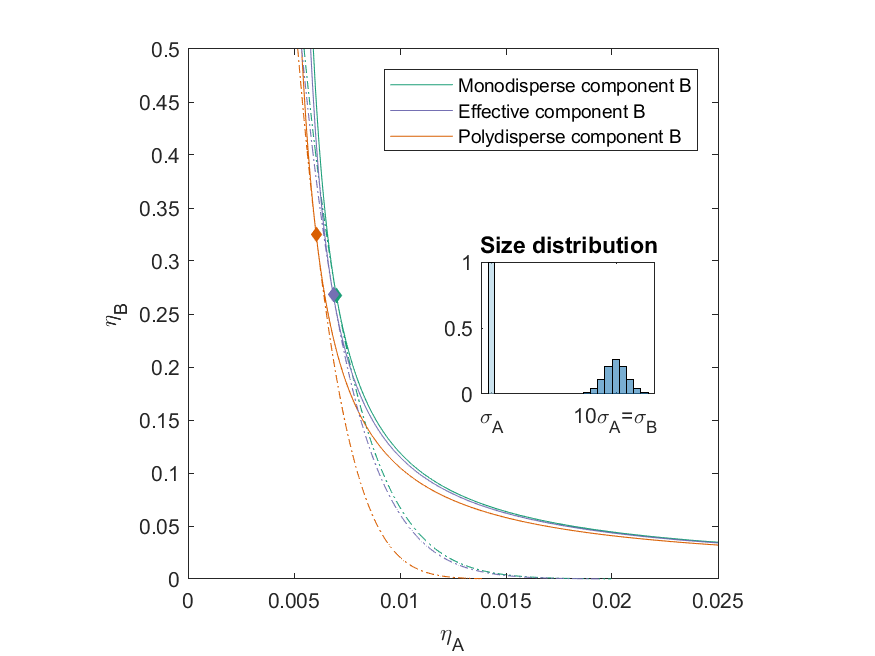}%
		}\hfill
		\subfloat[Left skewed distribution\label{fig:leftskew}]{%
			\includegraphics[trim=1cm 0cm 1cm 0cm, clip=true,width=0.42\textwidth]{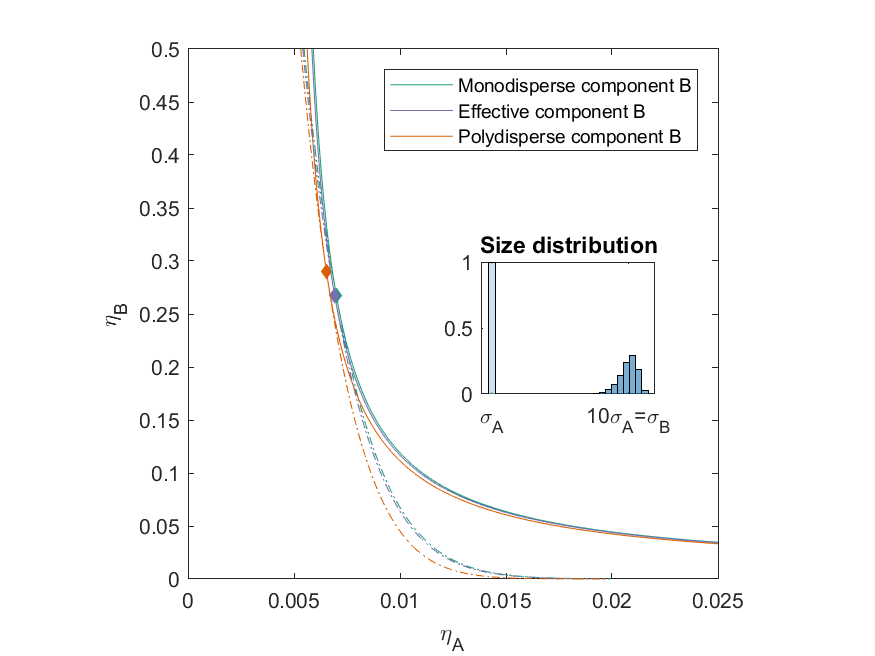}%
		}
		\subfloat[Right skewed distribution\label{fig:skewright}]{%
			\includegraphics[trim=1cm 0cm 1cm 0cm, clip=true,width=0.42\textwidth]{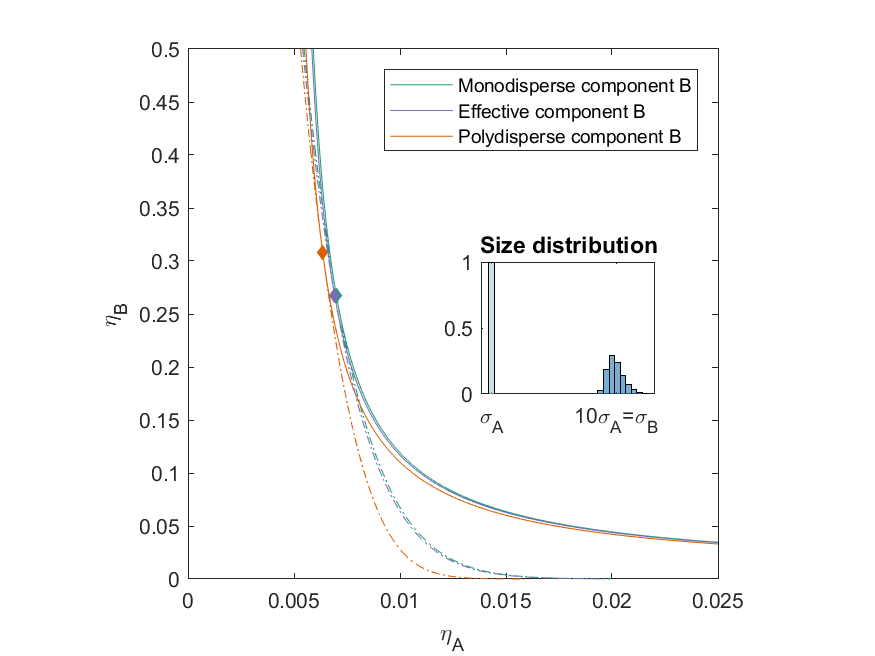}%
		}\hfill
		\subfloat[Bimodal distribution\label{fig:bimodal}]{%
			\includegraphics[trim=1cm 0cm 1cm 0cm, clip=true,width=0.42\textwidth]{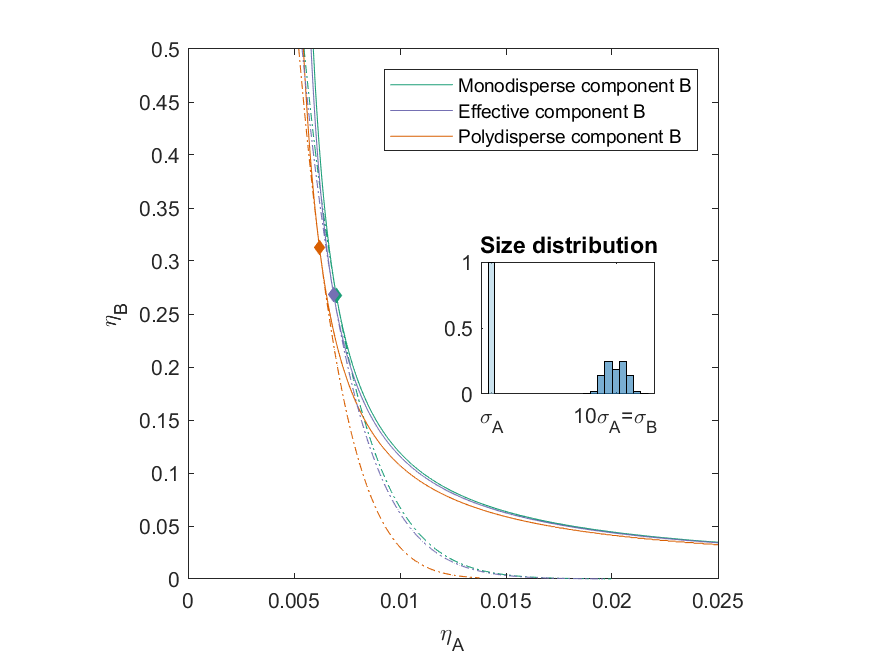}%
		}
		\hfill
		\caption{Phase diagram for binary (component $A$ and $B$) additive hard sphere mixture with size ratio $q = \sigma_A / \sigma_B =  1/10$, component $A$ is monodisperse, component $B$ is polydisperse (see size distributions for the different distribution), with a number average size 10 times the size of component $A$, plotted as a function of the partial packing fractions, $\eta_A$ and $\eta_B$. The spinodal (solid line) and binodal (dashed line) meet each other at the critical point (diamond)}
		\label{fig:9bins}
	\end{figure*}

	Now that we have a bit of an understanding of how polydispersity influences the critical point and the phase boundary, we increase the number of sub-components of $B$. The mixtures in figure \ref{fig:9bins} thus consist of ten components (one component $A$ and nine components $B$ in varying amounts and sizes, with different degrees of polydispersity, dependent on the considered distribution). We again calculated the phase diagram using both the simplified effective $2 \times 2$ virial coefficient matrix (the effective mixture) and the full $ 10 \times 10 $ virial coefficient matrix (the polydisperse mixture). Table \ref{tab:critical} gives the critical points for the polydisperse mixtures and table \ref{tab:fractionhist} allows for an easy comparison of the different distributions.

	\makeatletter\onecolumngrid@push\makeatother

	\clearpage
	\makeatletter\onecolumngrid@pop\makeatother
	
	\begin{widetext}
	
	\begin{table*}
		\caption{\label{tab:critical}Critical points for the different binary mixtures depending on the distribution of component $B$, see also figures \ref{fig:monodisperse} trough \ref{fig:9bins} and phase separated concentrations and volume fraction $\alpha$ of the different mixtures for specific parent concentration ($\eta_{A_{parent}} =0.010, \eta_{B_{parent}} =0.200$), depending on distribution of component $B$ see \ref{tab:fractionhist} for distribution of component $B$ in each phase.}
		\centering
		\begin{tabular}{l p{2.5 cm} p{3.5cm} p{3.5cm} c}
			\hline
			\textbf{Composition of component $B$} & \textbf{$\eta_{crit}$} &\textbf{Top phase}& \textbf{Bottom phase} & \textbf{$\alpha$}\\  
			\hline
			\ref{fig:monodisperse}: Monodisperse (PD: 0.00) & (0.007, 0.267) & $\eta$ (0.012,0.030) & $\eta$ (0.004,0.849)& 0.793\\
			& & PD: 0.00, Size: 1.00 & PD: 0.00, Size: 1.00 & \\
			\hline
			\ref{fig:PD5050smal}: Narrow 2 peaks (PD: 4.00) & (0.007, 0.280) & $\eta$ (0.011,0.034) & $\eta$ (0.004,0.881)& 0.804\\
			& & PD: 3.45, Size: 0.98 & PD: 3.97, Size: 1.00 & \\
			\hline
			\ref{fig:PD5050big}: Broad 2 peaks (PD: 8.00) & (0.006, 0.315) & $\eta$ (0.011,0.045) & $\eta$ (0.003,0.994)& 0.837\\
			& & PD: 4.86, Size: 0.93 & PD: 7.58, Size: 1.02 & \\
			\hline
			\ref{fig:PD2575}: Left skewed 2 peaks (PD: 6.93) & (0.006, 0.290) & $\eta$ (0.011,0.037) & $\eta$ (0.004,0.952)& 0.822\\
			& & PD: 7.92, Size: 0.93 & PD: 5.57, Size: 1.02 & \\
			\hline
			\ref{fig:PD7525}: Right skewed 2 peaks (PD: 6.93) & (0.006, 0.326)& $\eta$ (0.011,0.044) & $\eta$ (0.004,0.941)& 0.826\\
			& & PD: 2.36, Size: 0.96 & PD: 7.33, Size: 1.01 & \\
			\hline
			\ref{fig:PD9010}: Left skewed 2 peaks extreme (PD: 4.80) & (0.007, 0.276)  & $\eta$ (0.011,0.032) & $\eta$ (0.004,0.895)& 0.805\\
			& & PD: 8.12, Size: 0.96 & PD: 3.53, Size: 1.01 & \\
			\hline
			\ref{fig:PD1090}: Right skewed 2 peaks extreme (PD: 4.80) & (0.007, 0.304)  & $\eta$ (0.011,0.036) & $\eta$ (0.004,0.885)& 0.807\\
			& & PD: 1.20, Size: 0.98 & PD: 5.13, Size: 1.00 & \\
			\hline
			\ref{fig:smallgaus}: Narrow Gaussian (PD: 4.05) & (0.007, 0.281) & $\eta$ (0.011,0.034) & $\eta$ (0.004,0.881)& 0.804\\
			& & PD: 3.90, Size: 0.98 & PD: 3.96, Size: 1.00 & \\
			\hline
			\ref{fig:biggaus}: Broad Gaussian (PD: 8.11) & (0.006, 0.325) & $\eta$ (0.011,0.042) & $\eta$ (0.003,0.980)& 0.832\\
			& & PD: 7.14, Size: 0.93 & PD: 7.36, Size: 1.02 & \\
			\hline
			\ref{fig:leftskew}: Left skewed (PD: 6.02) & (0.007, 0.291) & $\eta$ (0.011,0.039) & $\eta$ (0.004,0.918)& 0.817\\
			& & PD: 4.33, Size: 0.96 & PD: 6.03, Size: 1.01 & \\
			\hline
			\ref{fig:skewright}: Right skewed (PD: 6.02) & (0.006, 0.308) & $\eta$ (0.011,0.036) & $\eta$ (0.004,0.921)& 0.815\\
			& & PD: 7.02, Size: 0.95 & PD: 5.29, Size: 1.01 & \\ 
			\hline
			\ref{fig:bimodal}: Bimodal (PD: 7.44) & (0.006, 0.313)& $\eta$ (0.011,0.042) & $\eta$ (0.004,0.969)& 0.830\\
			& & PD: 6.08, Size: 0.94 & PD: 6.94, Size: 1.02 & \\
			\hline
		\end{tabular} 
	\end{table*}	
	\end{widetext}

	From the figure and table we can conclude that, the standard deviation of the component size of the polydisperse component plays a big role in moving the critical point and phase boundary. The standard deviation for $B$ in figure \ref{fig:biggaus} is twice the standard deviation for $B$ in figure \ref{fig:smallgaus}. It is also clear that the type of distribution also plays a significant role in the concentration of the critical point and position of the phase boundary. The sizes in $B$ for figure \ref{fig:biggaus} and figure \ref{fig:bimodal} are the same, however, each size is present with a different frequency. The distribution in figure \ref{fig:biggaus} is Gaussian and the sizes in figure \ref{fig:bimodal} are bimodal, which means that the particles with sizes just larger and just smaller than the mean are present in a larger number. This causes to shift the critical point to slightly lower packing fraction of $B$ in \ref{fig:bimodal} compared to \ref{fig:biggaus}. The distribution of $B$ in mixture \ref{fig:PD5050smal} is comparable to the mixture in \ref{fig:smallgaus}, just with fewer sub-components. The position of the critical point is for both mixtures very comparable. In the same way is the distribution of $B$ in \ref{fig:PD5050big} comparable to the mixture in \ref{fig:biggaus} and \ref{fig:bimodal}. However, the increased number of sub-components results in a slightly higher and lower concentration of the $B$ component at the critical point.
	
	The distributions of $B$ in mixtures \ref{fig:leftskew} and \ref{fig:skewright} are skewed. In this regard are they comparable to the mixtures in \ref{fig:PDskew}. For these mixtures we see that the right skewed distribution also causes the critical point to move to higher concentrations of $B$ compared to the mixture with the left skewed distribution. 

	\subsection{\label{ss:Fraction}Fractionation of polydisperse component}
	
	\begin{table*}
		\caption{\label{tab:fractionhist}Phase separation of different mixtures and fractionation of component $B$ for specific parent distribution ($\eta_{A_{parent}} =0.010, \eta_{B_{parent}} =0.200$), depending on distribution of component $B$, see also figures \ref{fig:monodisperse} trough \ref{fig:9bins}}
		\centering
		\begin{tabular}{p{5 cm} c c c}
			\hline
			\multicolumn{1}{l}{\textbf{Composition of component $B$}}&
			\multicolumn{1}{c}{\textbf{Parent distribution}}&
			\multicolumn{1}{c}{\textbf{Top phase}}&
			\multicolumn{1}{c}{\textbf{Bottom phase}}\\  
			\hline
			\ref{fig:monodisperse}: Monodisperse
			&\adjustbox{valign=t}{\includegraphics[trim = 1.3cm .5cm 9.7cm 8.5cm,clip=true, width= 2.5 cm]{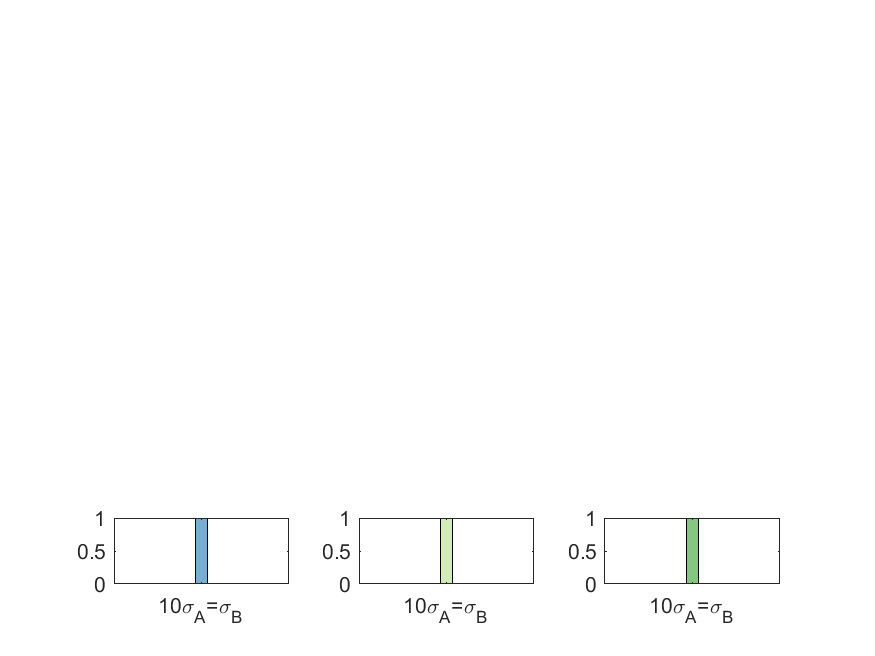}}  & \adjustbox{valign=t}{\includegraphics[trim = 5.3cm .5cm 5.7cm 8.5cm,clip=true, width= 2.5 cm]{PDmix_histograms_AB_25_i1_q10_delta0.png}} &\adjustbox{valign=t}{\includegraphics[trim = 9.5cm .5cm 1.5cm 8.5cm,clip=true, width= 2.5 cm]{PDmix_histograms_AB_25_i1_q10_delta0.png}} \\
			\ref{fig:PD5050smal}: Narrow 2 peak distribution 
			&\adjustbox{valign=t}{\includegraphics[trim = 1.3cm .5cm 9.7cm 8.5cm,clip=true, width= 2.5 cm]{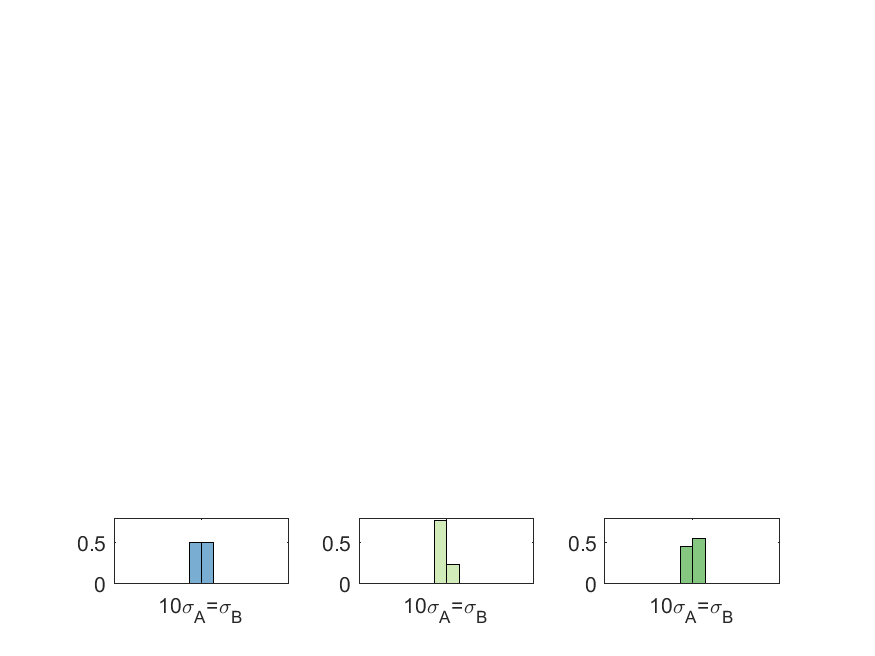}}  & \adjustbox{valign=t}{\includegraphics[trim = 5.3cm .5cm 5.7cm 8.5cm,clip=true, width= 2.5 cm]{PDmix_histograms_AB_25_i2_q10_delta0.png}} &\adjustbox{valign=t}{\includegraphics[trim = 9.5cm .5cm 1.5cm 8.5cm,clip=true, width= 2.5 cm]{PDmix_histograms_AB_25_i2_q10_delta0.png}} \\
			\ref{fig:PD5050big}: Broad 2 peak distribution 
			&\adjustbox{valign=t}{\includegraphics[trim = 1.3cm .5cm 9.7cm 8.5cm,clip=true, width= 2.5 cm]{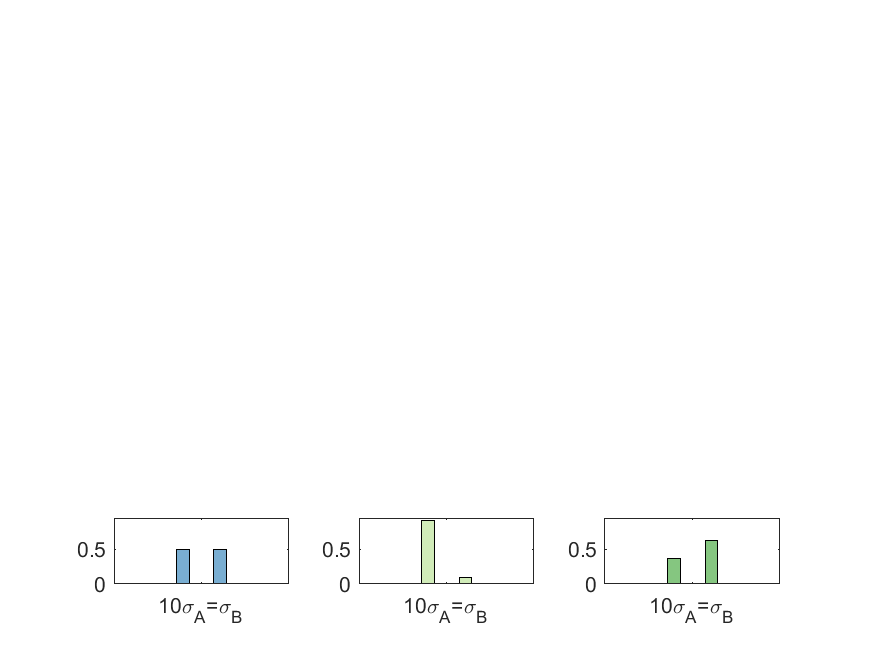}}  & \adjustbox{valign=t}{\includegraphics[trim = 5.3cm .5cm 5.7cm 8.5cm,clip=true, width= 2.5 cm]{PDmix_histograms_AB_25_i3_q10_delta0.png}} &\adjustbox{valign=t}{\includegraphics[trim = 9.5cm .5cm 1.5cm 8.5cm,clip=true, width= 2.5 cm]{PDmix_histograms_AB_25_i3_q10_delta0.png}} \\
			\ref{fig:PD2575}: Left skewed 2 peak distribution 
			&\adjustbox{valign=t}{\includegraphics[trim = 1.3cm .5cm 9.7cm 8.5cm,clip=true, width= 2.5 cm]{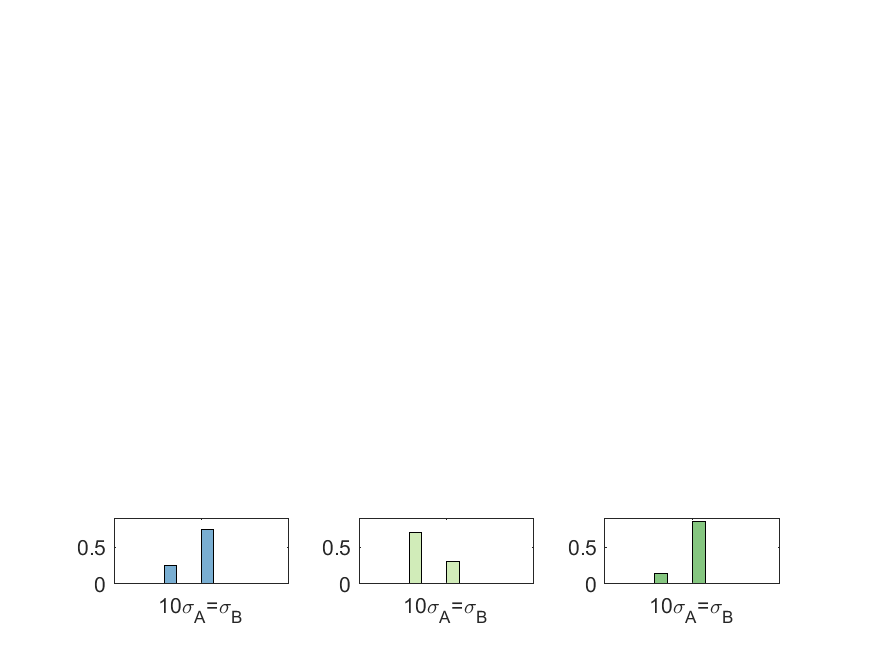}}  & \adjustbox{valign=t}{\includegraphics[trim = 5.3cm .5cm 5.7cm 8.5cm,clip=true, width= 2.5 cm]{PDmix_histograms_AB_25_i4_q10_delta0.png}} &\adjustbox{valign=t}{\includegraphics[trim = 9.5cm .5cm 1.5cm 8.5cm,clip=true, width= 2.5 cm]{PDmix_histograms_AB_25_i4_q10_delta0.png}} \\
			\ref{fig:PD7525}: Right skewed 2 peak distribution 
			&\adjustbox{valign=t}{\includegraphics[trim = 1.3cm .5cm 9.7cm 8.5cm,clip=true, width= 2.5 cm]{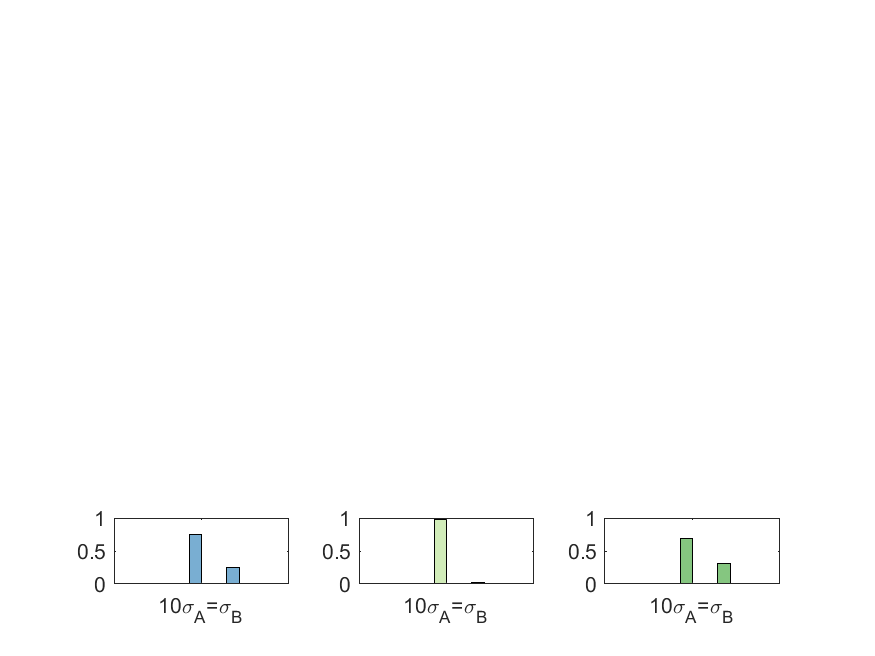}}  & \adjustbox{valign=t}{\includegraphics[trim = 5.3cm .5cm 5.7cm 8.5cm,clip=true, width= 2.5 cm]{PDmix_histograms_AB_25_i5_q10_delta0.png}} &\adjustbox{valign=t}{\includegraphics[trim = 9.5cm .5cm 1.5cm 8.5cm,clip=true, width= 2.5 cm]{PDmix_histograms_AB_25_i5_q10_delta0.png}}  \\
			\ref{fig:PD9010}: Left skewed 2 peak distribution extreme 
			&\adjustbox{valign=t}{\includegraphics[trim = 1.3cm .5cm 9.7cm 8.5cm,clip=true, width= 2.5 cm]{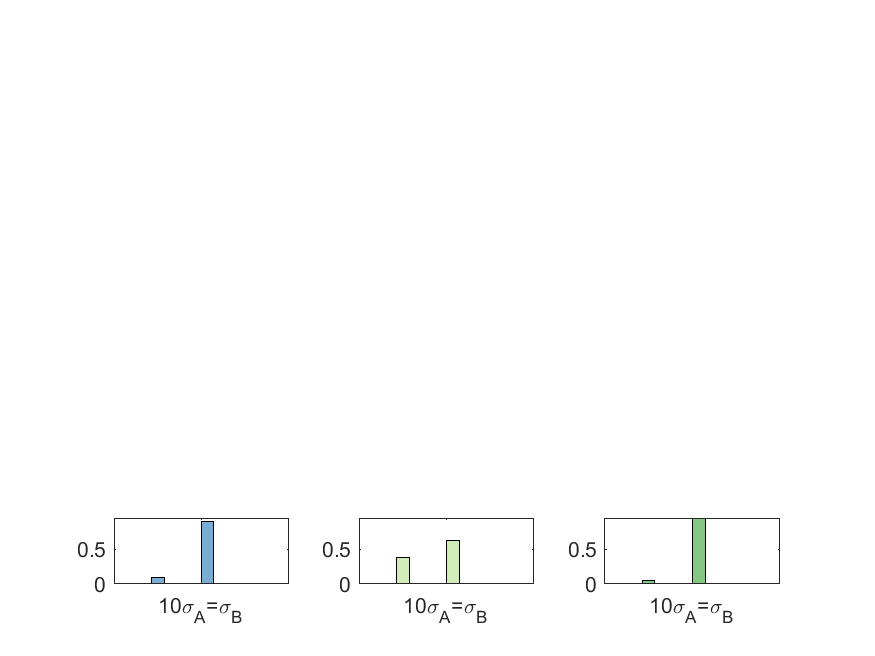}}  & \adjustbox{valign=t}{\includegraphics[trim = 5.3cm .5cm 5.7cm 8.5cm,clip=true, width= 2.5 cm]{PDmix_histograms_AB_25_i6_q10_delta0.png}} &\adjustbox{valign=t}{\includegraphics[trim = 9.5cm .5cm 1.5cm 8.5cm,clip=true, width= 2.5 cm]{PDmix_histograms_AB_25_i6_q10_delta0.png}} \\
			\ref{fig:PD1090}: Right skewed 2 peak distribution extreme 
			&\adjustbox{valign=t}{\includegraphics[trim = 1.3cm .5cm 9.7cm 8.5cm,clip=true, width= 2.5 cm]{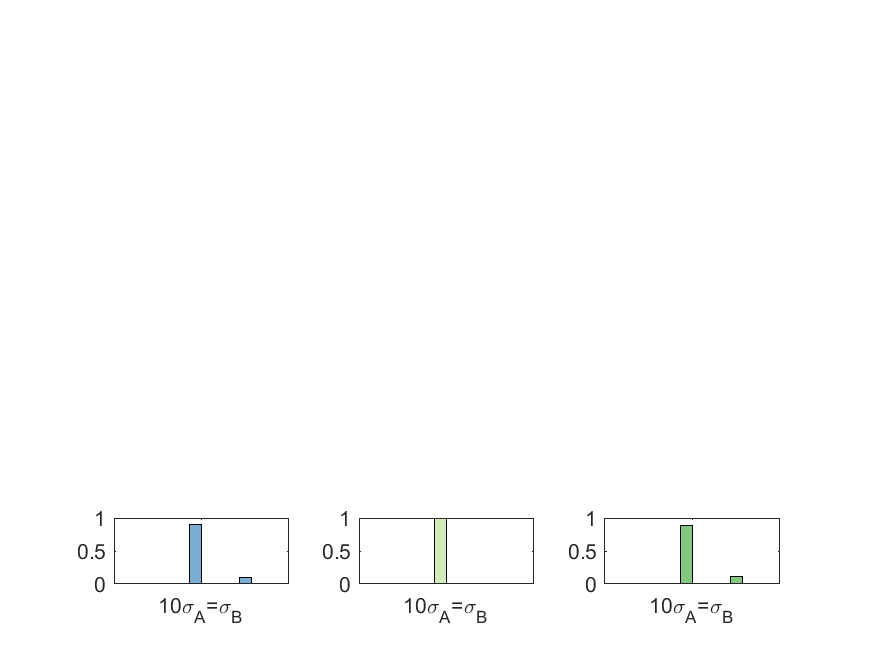}}  & \adjustbox{valign=t}{\includegraphics[trim = 5.3cm .5cm 5.7cm 8.5cm,clip=true, width= 2.5 cm]{PDmix_histograms_AB_25_i7_q10_delta0.png}} &\adjustbox{valign=t}{\includegraphics[trim = 9.5cm .5cm 1.5cm 8.5cm,clip=true, width= 2.5 cm]{PDmix_histograms_AB_25_i7_q10_delta0.png}} \\
			\ref{fig:smallgaus}: Narrow Gaussian distribution  
			&\adjustbox{valign=t}{\includegraphics[trim = 1.3cm .5cm 9.7cm 8.5cm,clip=true, width= 2.5 cm]{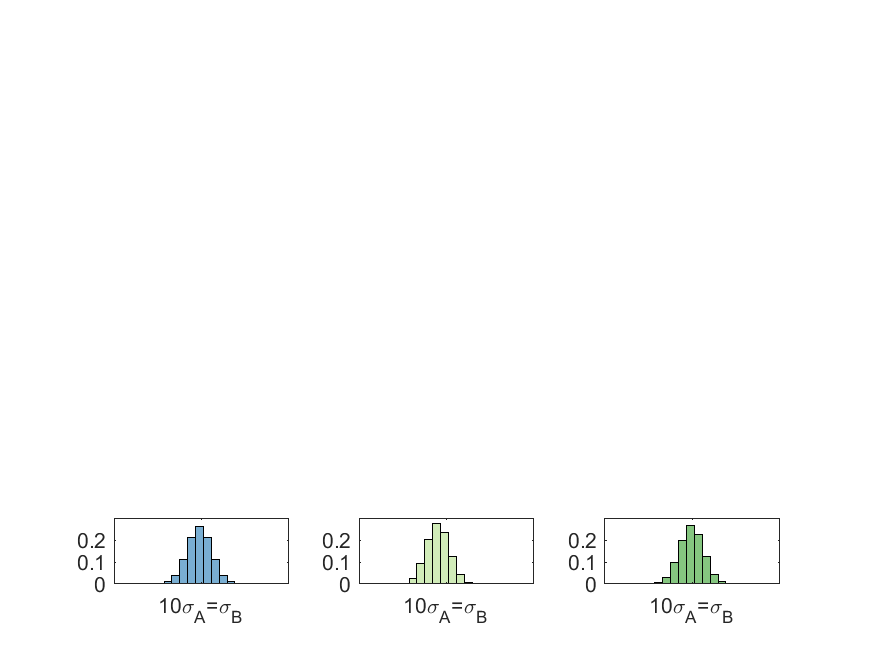}}  & \adjustbox{valign=t}{\includegraphics[trim = 5.3cm .5cm 5.7cm 8.5cm,clip=true, width= 2.5 cm]{PDmix_histograms_AB_25_i8_q10_delta0.png}} &\adjustbox{valign=t}{\includegraphics[trim = 9.5cm .5cm 1.5cm 8.5cm,clip=true, width= 2.5 cm]{PDmix_histograms_AB_25_i8_q10_delta0.png}} \\ 
			\ref{fig:biggaus}: Broad Gaussian distribution  
			&\adjustbox{valign=t}{\includegraphics[trim = 1.3cm .5cm 9.7cm 8.5cm,clip=true, width= 2.5 cm]{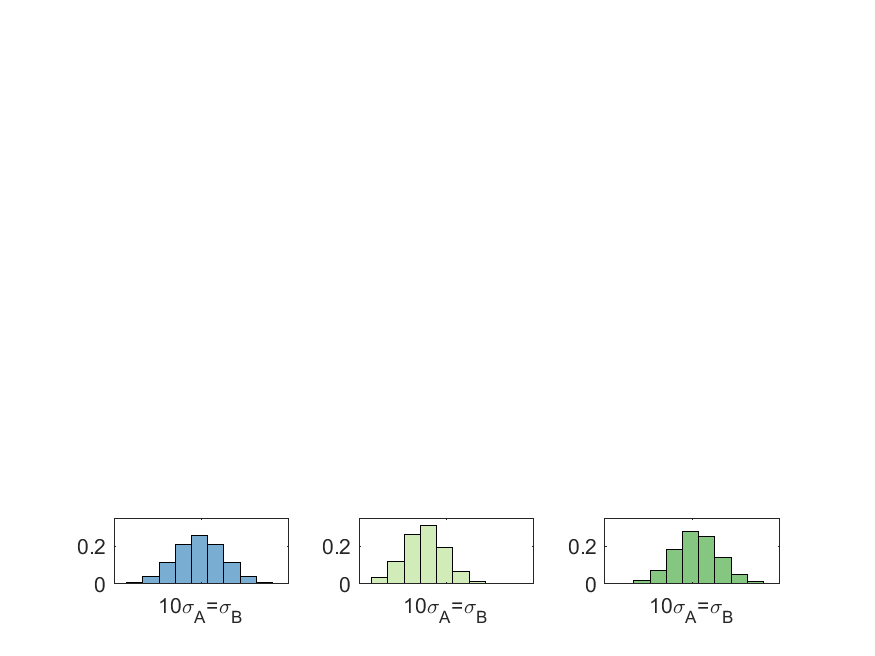}}  & \adjustbox{valign=t}{\includegraphics[trim = 5.3cm .5cm 5.7cm 8.5cm,clip=true, width= 2.5 cm]{PDmix_histograms_AB_25_i9_q10_delta0.png}} &\adjustbox{valign=t}{\includegraphics[trim = 9.5cm .5cm 1.5cm 8.5cm,clip=true, width= 2.5 cm]{PDmix_histograms_AB_25_i9_q10_delta0.png}} \\
			\ref{fig:leftskew}: Left skewed distribution 
			&\adjustbox{valign=t}{\includegraphics[trim = 1.3cm .5cm 9.7cm 8.5cm,clip=true, width= 2.5 cm]{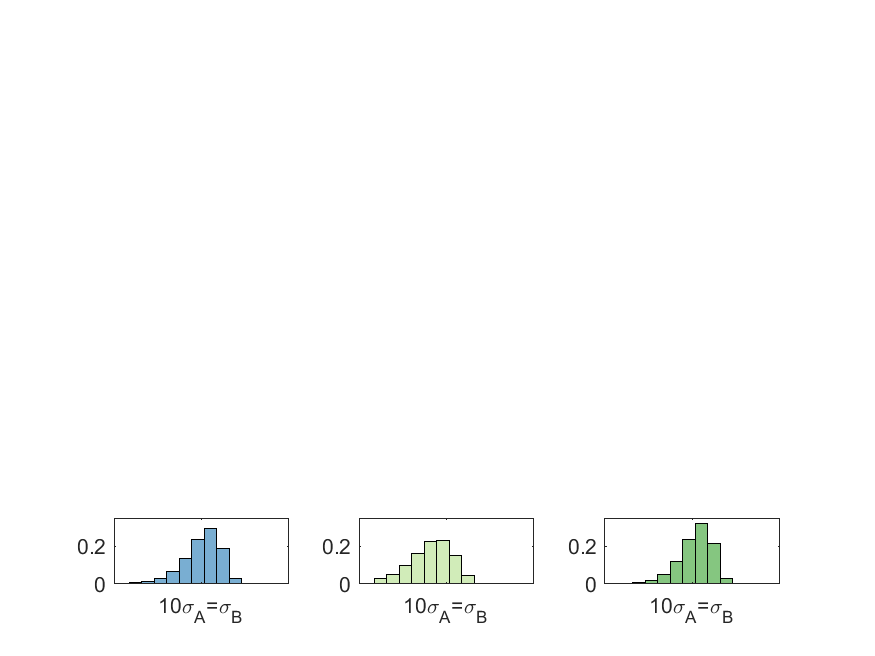}}  & \adjustbox{valign=t}{\includegraphics[trim = 5.3cm .5cm 5.7cm 8.5cm,clip=true, width= 2.5 cm]{PDmix_histograms_AB_25_i12_q10_delta0.png}} &\adjustbox{valign=t}{\includegraphics[trim = 9.5cm .5cm 1.5cm 8.5cm,clip=true, width= 2.5 cm]{PDmix_histograms_AB_25_i12_q10_delta0.png}} \\
			\ref{fig:skewright}: Right skewed distribution 
			&\adjustbox{valign=t}{\includegraphics[trim = 1.3cm .5cm 9.7cm 8.5cm,clip=true, width= 2.5 cm]{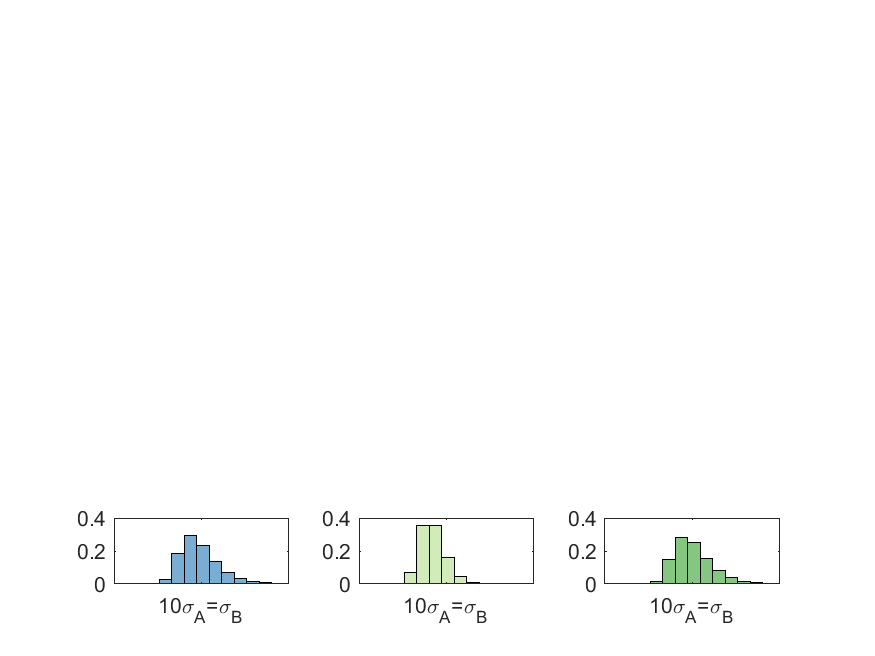}}  & \adjustbox{valign=t}{\includegraphics[trim = 5.3cm .5cm 5.7cm 8.5cm,clip=true, width= 2.5 cm]{PDmix_histograms_AB_25_i11_q10_delta0.png}} &\adjustbox{valign=t}{\includegraphics[trim = 9.5cm .5cm 1.5cm 8.5cm,clip=true, width= 2.5 cm]{PDmix_histograms_AB_25_i11_q10_delta0.png}} \\
			\ref{fig:bimodal}: Bimodal distribution 
			&\adjustbox{valign=t}{\includegraphics[trim = 1.3cm .5cm 9.7cm 8.5cm,clip=true, width= 2.5 cm]{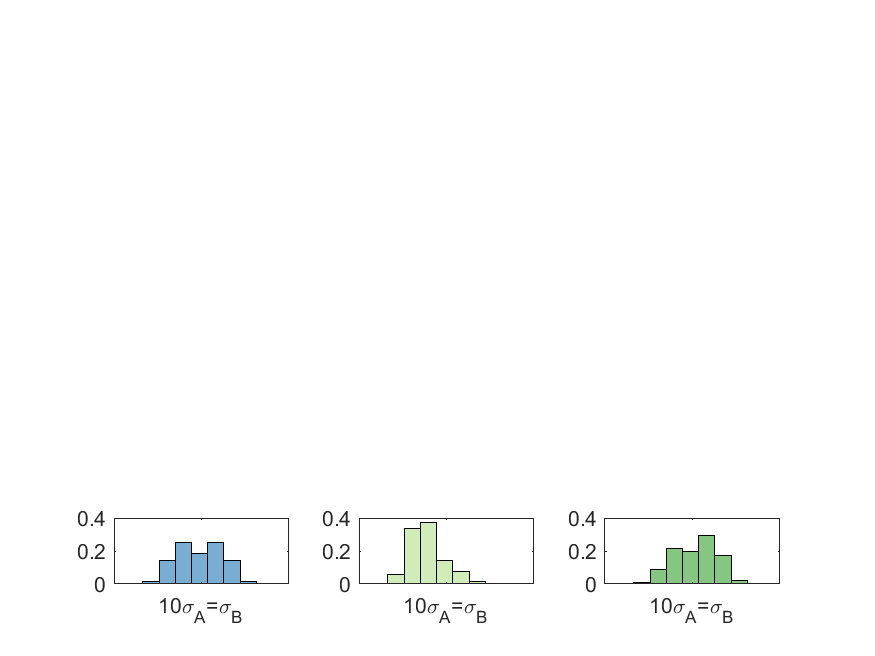}}  & \adjustbox{valign=t}{\includegraphics[trim = 5.3cm .5cm 5.7cm 8.5cm,clip=true, width= 2.5 cm]{PDmix_histograms_AB_25_i10_q10_delta0.png}} &\adjustbox{valign=t}{\includegraphics[trim = 9.5cm .5cm 1.5cm 8.5cm,clip=true, width= 2.5 cm]{PDmix_histograms_AB_25_i10_q10_delta0.png}}  \\
			\hline
		\end{tabular} 
	\end{table*}	

	Upon phase separation, particles will move to a preferential phase in order to minimize the Helmholtz free energy. One phase is enriched in component $A$, whilst the other is enriched in component $B$. Even though each phase is enriched in one component, the other component is still present in lower concentrations. We investigated the phase separation of a specific parent mixture ($\eta_{A_{parent}} =0.010, \eta_{B_{parent}} =0.200$) for the different distributions of component $B$, in terms of volume fraction of both components in each phase, degree of polydispersity of component $B$, average size of component $B$ in child phase compared to the average size of component $B$ in the parent phase and the volumefraction of the phases ($\alpha$), see table \ref{tab:fractionhist}. 
	
	Polydispersity in the sizes of component $B$ in the parent phase causes significant fractionation of component $B$ in the child phases. The phase enriched in component $A$, the smaller component, contains also relatively more of the smaller components of $B$ than the phase enriched in component $B$. The size ratio of the average size of component $B$ compared to the average size of component $B$ in the parent phase is smaller than 1. The polydispersity of component $B$ influences also both the composition in each phase as well as the volume fraction of the phases. The volume fraction of the top phase, the phase enriched in component $A$ (phase with lowest volume fraction in components) increases with polydipsersity. The bottom phase has a higher volume fraction of component $B$ with higher polydispersity. In general, but most pronounced for the mixture in \ref{fig:PD2575}, we observe that the smaller sub-components favor the top phase (the phase enriched in the small particles $A$), while the larger sub-components favor the bottom phase (the phase enriched in the larger particles $B$). We like to note that these observations as obtained from the virial approach are in line with previous theoretical work using the UNIQUAC model by Kang and Sandler \cite{Kang1988} and the Florry Huggins theory \cite{VanHeukelum2003} and also experimental work \cite{Edelman2001} \cite{VanderKooij2000} \cite{Loret2005}. Evans and co-authors also note that the skewness of the polydisperse parent distribution plays an important role in the fractionation \cite{Evans1998}. Also Paricaud \cite{Paricaud2008} found that the largest polydipserse particle favored the phase poor in the smallest particle. This follows from the relatively increased size incompatibility between the smallest particle and the largest particle in the system.
	
	Depending on the type of distribution, phase separation causes the degree of polydispersity to decrease. This is in both phases for all mixtures with a symmetric distribution of $B$ in the parent phase. For these mixtures, the degree of polydispersity is ati its lowest for the phase enriched in $A$.
	
	\section{Conclusion}
	We find that the largest species in the polydisperse component causes the largest shift in the position of the phase boundary, critical point and spinodal compared to the binary monodisperse binary mixtures. Upon phase separation, the polydisperse component fractionates. The smaller species of the polydisperse component favor the phase enriched in the small component, while the larger species remain in the phase enriched in the larger component. The top phase, the phase enriched in the small component, has a larger volume and this volume increases with polydispersity.	
	The virial approach we used yields results in line with previous theoretical and experimental work on polydisperse mixtures, and at the same time allows for direct experimental testing using virial coefficients obtained from membrane osmometry.
	\begin{acknowledgments}
		We would like to thank Paul Venema and Arjen Bot for their interest in the work and the helpful discussions.
	\end{acknowledgments}

	\printnomenclature
	\bibliographystyle{unsrt} 
	\bibliography{PolydispersityLit}
\end{document}